\documentclass[a4paper,10pt]{article}
\usepackage{amsfonts}
\usepackage{amssymb}
\usepackage{amsmath}
\usepackage{amsthm}
\usepackage{latexsym}
\usepackage{layout}
\usepackage[english]{babel}

\usepackage{epsfig}
\usepackage{graphicx}
\usepackage{pdfsync}

\numberwithin{equation}{section} 
\newtheorem{theorem}{Theorem}[section]
\newtheorem{proposition}[theorem]{Proposition}
\newtheorem{lemma}[theorem]{Lemma}
\newtheorem{remark}[theorem]{Remark}
\newtheorem{definition}[theorem]{Definition}

\newtheorem{example}[theorem]{Example}

\newcommand{\hide}[1]   {}

\newcommand{\eps}       {\varepsilon}

\renewcommand{\H}        {{\mathcal H}}

\newcommand{\R}         {\mathbb{R}}

\newcommand{\nada}[1]    {}

\newcommand{\Z}{\mathbb Z}
\newcommand{\ZZ}{\mathbb Z}
\newcommand{\N}{\mathbb N}
\newcommand{\RR}{\mathbb R}

\def\A{{\mathcal A}}

\def\e{\eps}
\def\d{\delta}

\begin{document}

\title{Interfaces, modulated phases and textures in lattice systems}

\author{
Andrea Braides and
Marco Cicalese 
}
\date{}
\maketitle

\begin{abstract}\noindent We introduce a class of $n$-dimensional (possibly inhomogeneous) spin-like lattice systems presenting modulated phases with possibly different textures. Such systems can be 
parameterized according to the number of ground states, and can be described by 
a phase-transition energy which we compute by means of variational techniques. 
Degeneracies due to frustration are also discussed.
\end{abstract}

\section{Introduction}\label{secint}
The object of the analysis in this work is the study of variational scalar spin-like lattice systems 
in which local minimization gives rise to regular patterns, and at the same time geometric incompatibilities (e.g., those imposed by boundary conditions) force the appearance of interfaces between patterns or between variants of the same pattern differing by a lattice translation. 

Physical systems exhibiting ground states showing such features have been discussed by Seul and Andelman in the celebrated paper \cite{SA}. Those authors propose frustration as a possible common mechanism explaining the emergence of these periodic structures, also known as {\it spatially modulated phases}.  
The frustration is due to the presence of competing interactions leading to the impossibility of minimizing all the interactions at the same time. Two paradigmatic examples in the framework of spin systems are the Ising model with short-range ferromagnetic and long-range antiferromagnetic interactions studied by variational methods by Giuliani {\em et al.}~\cite{GLL, GLL3, GLS} and the microemulsion models of the type introduced by Ciach {\em et al.}~\cite{CHS}, where the frustration mechanism is instead due to only competing short-range interactions. Both of them have well-known continuous analogues as for instance the Ohta-Kawasaki model for diblock copolymers \cite{OK} and the Coleman-Mizel model for second order materials \cite{CMM}. 

The results in this paper are part of a general analysis of interfacial energies for lattice systems by variational methods, for which an overview can be found in \cite{ABC-libro} (see also \cite{ICM}).
In that context, many of the results for energies depending on scalar spin variables concern ferromagnetic interactions \cite{AlGe,B-planelike,BPiatn,CDLL,BCPS}, so that they can be approximated by continuum energies with one or more parameters taking only the values $\pm1$. In \cite{AlGe} and \cite{BPiatn} mixtures of ferromagnetic and antiferromagnetic interactions 
are also considered, provided their ground states are the constant ones, or that
antiferromagnetic interactions influence
the form of the ground state only in separate disconnected zones, respectively.
In the second case the relevant parameter is the constant value in the connected component
of the lattice ({\em majority phase}), still taking only two values. In \cite{BPiatn} dilute systems have been considered,
where some of the interaction coefficients are zero, so that the value of the variable $u$ is not determined on some nodes, but again the relevant parameters are still the majority phases
if the coefficients are non-zero in an infinite connected region. The possibility
of the existence of non-trivial oscillating ground states have been observed in \cite{ABC}, where the case of antiferromagnetic nearest and next-to-nearest neighbor interactions are examined in detail (see also Example \ref{ABC_ex} below). In that case the number of relevant parameters is four and cannot be reduced to the ferromagnetic
description.  That paper has been the starting point of our analysis.

We consider discrete systems depending on a variable $u$ defined on the nodes of a lattice $\mathcal L$  in $\RR^n$ and 
taking a finite number of values. In particular, we have  in mind the case of scalar spin systems where the variable only takes the values $+1$ and $-1$. On such a system we can consider an energy depending in principle on the interactions between all points in the lattice, which we can write, in the assumption of  invariance by translations of the lattice, as
\begin{equation}\label{ef}
F(u)=\sum_i \phi(\{u^{i+j}\}_j),
\end{equation}
where the index $i$ runs on all values in (a portion of) the lattice $\mathcal L$ and $u^i$ is the value of $u$ at $i$.
In the simplest case of spin systems in ${\mathcal L}=\ZZ^n$, we can consider as a model
an energy density
$$
\phi(\{u^{j}\}_j)= -\sum_{j\in \ZZ^n\setminus\{0\}}\sigma_{j}\, u^ju^0
$$
representing the sum of the pair interactions of each spin at the site $j$ with the one at the origin.
The energy $F$ is  then obtained by summing up the contributions centered at  all $i\in\ZZ^n$.
The interaction coefficients $\sigma_{j}$ may have different signs, and in particular be negative
({\em antiferromagnetic interactions}) thus favoring oscillating $u$.

As an example we consider a two-dimensional system parameterized on $\ZZ^2$ with dominating antiferromagnetic interactions between points at distance two; i.e., we can suppose
that
\begin{equation}\label{NN-NNN}
\phi(\{u^{j}\}_j)= u^{(2,0)}u^{(0,0)}+  u^{(0,2)}u^{(0,0)}+ u^{(-2,0)}u^{(0,0)}+  u^{(0,-2)}u^{(0,0)}
\end{equation}
$+$ nearest-neighbor terms which do not change the shape of minimizers of $\phi$. 
 \begin{figure}[h!]
\centerline{\includegraphics [width=3.5in]{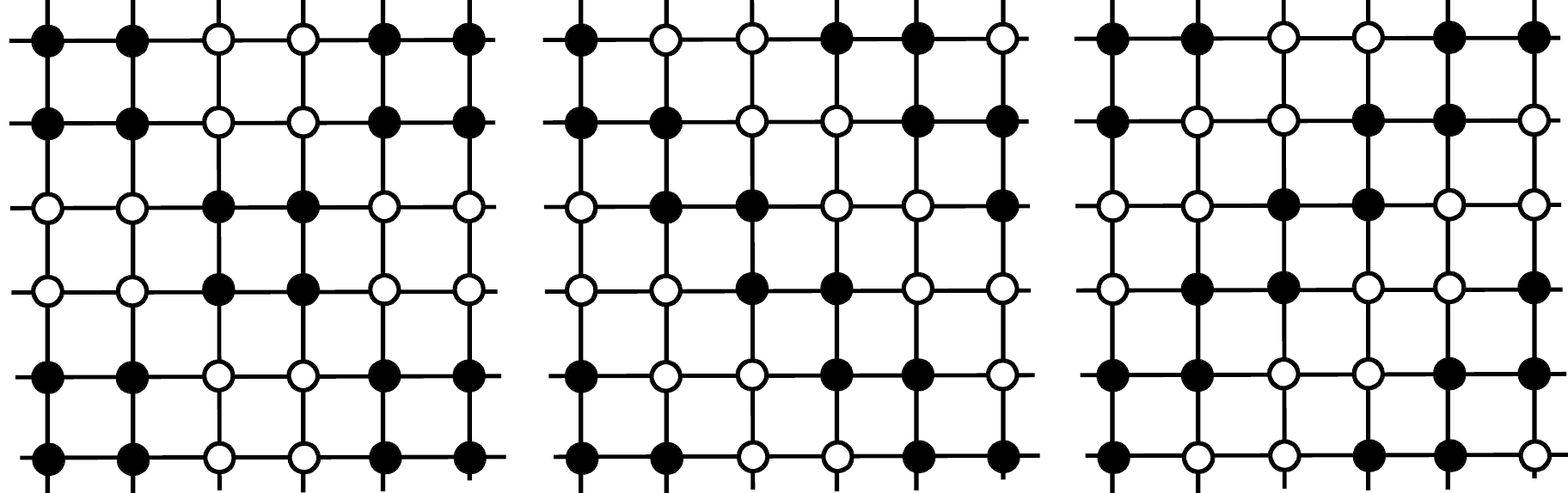}}
\caption{three textures of ground states}\label{patterns3}
   \end{figure}

In this case the minimization
of $\phi(\{u^{j}\}_j)$ at all $i$ leads to three types of {\em textures}, which turn out to be periodic
of period four in both directions, and which are represented in Fig.~\ref{patterns3} (black and white circles represent $+1$ and $-1$ spins, respectively).
For each pattern also the variants differing by a translation are minimizers. Such variants must be considered as different phases, even though they have the same texture. In this case we have eight spatially modulated phases for the textures with a checkerboard pattern and four each for the two striped patterns. The details are included in Example \ref{mpapb}.

In this first case the pointwise minimization of $\phi$ provides a description of all {\em ground states}. This might not be the case when the combination of the interactions forces {\em frustration}; i.e., there are no ground states minimizing all interactions at once. 
The simplest example  of this situation is that of 
 nearest and next-to-nearest neighbor antiferromagnetic interactions in $\ZZ^2$. 
 In that case, it is still possible to compute the (two-periodic) ground states after rewriting the
 energy density $\phi$ as depending on the four values at the vertices of a lattice square;
 namely (supposing that $\sigma_j$ depends only on the order of the neighbor),
\begin{eqnarray*}
\phi(\{u^{j}\}_j)&=& {1\over 2}\sigma_{\rm n}\Bigl(
u^{(1,0)}u^{(0,0)}+  u^{(1,1)}u^{(1,0)}+ u^{(0,1)}u^{(1,1)}+  u^{(0,0)}u^{(0,1)}\Bigr)
\\
&&
+ \sigma_{\rm nn}\Bigl(u^{(0,0)}u^{(1,1)}+ u^{(0,1)}u^{(1,0)}\Bigr).
\end{eqnarray*}
In this case, depending on the ratio $\sigma_{\rm n}/\sigma_{\rm nn}$, either we have two ground states with only one pattern of alternating spins, or 
we have two patterns of horizontal and vertical stripes for a total of four ground states (see Example \ref{ABC_ex}).

Note that even for very simple energy densities (e.g., nearest-neighbor antiferromagnetic interactions in the triangular lattice) we may have non-periodic minimal states (see also the examples in Section \ref{inho}). 

In this paper we explore the behavior of the energy $F$ in the case when minimal states are indeed periodic and in an energy regime when interfaces between such states are relevant. Such interfaces appear between textures with different patterns and also between two different modulated phases
corresponding to the same pattern, in which case we sometime use the terminology of {\em anti-phase} boundaries.

In order to describe the behavior of the energy $F$ we use a {\em discrete-to-continuum} approach as in \cite{ABC}: we first scale the energies as 
$$
F_\e(u)=\sum_i \e^{n-1}\phi(\{u^{i+j}\}_j),
$$
where now the functions $u$ are considered as describing the behavior on the scaled lattice
$\e\cal L$ (even though it will be notationally convenient to keep them parameterized on the original lattice $\cal L$). We then define a convergence of discrete functions $u_\e$ for which 
the energies $F_\e$ are equi-coercive. This is a fundamental step, for which we have to assume
that we can write $\phi$ in such a way that it is minimized (and is $0$ for computational convenience) exactly
on a finite family $v_1,\ldots, v_K$ of ground states. Indeed, a careful definition of ground states
is at the core of the present analysis. 

At this point our approach differs from others in the physical literature where it is customary to study even quite complex spin systems by mapping energetically admissible fields to a superposition of Ising-type variables (spin systems whose interactions can be described by an energy of Ising-type). If on one hand that approach turns out to be an efficient method in order to describe some general properties of the ground states, on the other hand it may oversimplify the description of transitions between different modulated phases as it may not capture the optimal geometry of the spin field at the interface.  Note, for example, that for the energies as in \eqref{ef} with potential given by \eqref{NN-NNN}, the optimal geometry at the interface strongly depends on the choice of the nearest neighbor interactions and hence influences the Wulff shapes of the system without changing the patterns of the ground-states.

In our setting we also assume that ground states cannot be mixed with arbitrarily low energy; this must be assumed and does not follow from the definition of ground states due to the very general form of $\phi$.
Under these hypotheses we can describe the behavior of sequences with equibounded energy as follows.
For each value of the label $l\in\{1,\ldots,K\}$ indexing the ground states we can find a set $A_l$ such that (upon the extraction of a subsequence) the functions $u_\e$ are equal to $v_l$ on ${1\over\e} A_l$ up to an asymptotically negligible set; i.e., the set $A_l$ describes approximately the location where the microscopic pattern of the functions $ u_\e$ on the lattice $\e\cal L$ is equal to $v_l$. 
In this way, we define a convergence $ u_\e\to {\mathbb A}:=(A_1,\ldots, A_K)$. We show that the sets $A_l$ form a partition of $\RR^n$ so that the description of the behavior of $u_\e$ is completely provided by $\mathbb A$.  Note that the (number of) {\em ground states themselves determine the parameterization} of this asymptotic description.
We also show that each such $A_l$ is a set of finite perimeter, which allows to give a notion of interface between phases with an orientation described by a normal vector $\nu$.
At this point, we can use the terminology of $\Gamma$-convergence (with respect to the convergence $ u_\e\to {\mathbb A})$ to describe the behavior of the energies $F_\e$.
In order to obtain a description in terms of interfacial energies we have to suppose some decay estimates on the interactions. Due to the very general form of the dependence of $\phi$ on $u$ 
this is also a delicate point. The decay condition that we assume formalizes the requirement that a 
perturbation at a single point of a ground state must have a finite energy (which then turns out to be negligible after scaling by $\e^{n-1}$). Under these assumptions, we prove that the $\Gamma$-limit takes the form of an interfacial energy between the elements of the partition, which can be written as 
$$
{\cal F}({\mathbb A}) =\sum_{l,l'=1}^K\int_{\partial A_l\cap \partial A_{l'}}\varphi_{l l'}(\nu_{ll'})\,d\H^{n-1},
$$
where $\nu_{ll'}$ denotes the exterior normal to $A_{l}$ on $\partial A_l\cap \partial A_{l'}$, and $\varphi_{l l'}$ is a norm, which is of {\em crystalline type} if the range of interactions is finite. This norm can be characterized by suitable homogenization formulas. Furthermore $(l,l',\nu)\mapsto \varphi_{l l'}(\nu)$ is $BV$-elliptic (see \cite{AmBrII}), and in particular $(l,l')\mapsto \varphi_{l l'}(\nu)$ is subadditive for fixed $\nu$. 

The same procedure can be repeated if the energy density depends on the site $i$; i.e., we have $\phi_i$ in place of $\phi$ in the definition of $F$, if $i\mapsto \phi_i$ is periodic. In this way we may include in our approach also energies describing periodic mixtures of spin energies and non-Bravais lattices. In our analysis we do not include random mixtures or random lattices, for an extension to which we refer to \cite{BPiatn,BPiatn2} and \cite{ACR}, respectively. We also mention that in some cases the decay requirements on the discrete energies can be relaxed, at the expenses of the appearance of non-local terms in the $\Gamma$-limit (see \cite{AlGe} for systems with ferromagnetic interactions). We further notice that the emergence of finitely many ground states can also be proved for some systems of ferromagnetic chiral molecules in which the domain of the discrete energy is subject to a global geometric constraint (see \cite{BGP}).

The proof of the convergence theorem requires several technicalities due to the general form
of the dependence of the energy densities; in particular all arguments involving the locality of the
limit energies must be worked out using only the decay properties of the energies. Another difficulty is due to the relative lack of results for partitions into sets of finite perimeter, in particular
strong density results of smooth partitions. This forces to use arguments where only estimates
on the $\H^{n-1}$ measure of partitions are necessary.


\section{Notation and Preliminaries}\label{Sect:preliminari}

Given $n\in\N$ we denote by $\{e_1,e_2,\dots,e_n\}$ the standard basis in $\R^n$. For any $x,y\in\RR^n$ we denote by $(x,y)$ the usual scalar product in $\RR^n$ and define $|x|=\sqrt{(x,x)}$. We also denote by $\|\cdot\|_{\infty}$ the $l_{\infty}$-norm in $\RR^n$ defined as $\|x\|_{\infty}=\max\{|x_i|,\ i\in\{1,2,\dots,n\}\}$ and $\|\cdot\|_{1}$ the $l_{1}$-norm in $\RR^n$ defined as $\|x\|_{1}=|x_1|+\cdots+|x_n|$. Then, for any set $A\subset\R^n$ we define the distances
\begin{eqnarray*}
{\rm dist}_\infty(x, A)&:=&\inf\{\|x-y\|_\infty,\ y\in A\},\\
{\rm dist}(x, A)&:=&\inf\{|x-y|,\ y\in A\}.
\end{eqnarray*}  
For all $t\in\R$ the symbol $\lfloor t\rfloor$ denotes the integer part of $t$.
Given $x=(x_{1},\dots,x_{n})\in\R^{n}$ we denote by $\lfloor x\rfloor$ the vector 
$(\lfloor x_{1}\rfloor,\dots, \lfloor x_{n}\rfloor)$.
 For $r>0$, by $B_{r}$ we denote the $n$-dimensional ball of radius $r$ centered at the origin. By ${\mathcal S^{n-1}}$ we denote the boundary of $B_{1}$; that is, the set of all unitary vectors in $\R^n$. Given $\nu\in {\mathcal S^{n-1}}$ we define $$\Pi_\nu:=\{x\in\R^n:\ (x,\nu)=0\}$$ the hyperplane orthogonal to $\nu$ and denote by $Q_\nu$ a given $n$-dimensional open cube centered at the origin having one face orthogonal to $\nu$ and side-length $1$. When $\nu=e_i$ for some $i\in\{1,\dots,n\}$, we choose $Q_{e_i}=Q:=(-\frac{1}{2},\frac{1}{2})^n$.  
 

For any measurable set $A\subset\RR^n$ we denote by $|A|$ the $n$-dimensional Lebesgue measure of $A$. We denote by ${\cal H}^{n-1}$ the $n-1$-dimensional Hausdorff measure.\\

Given $m\in\N$, $X\subseteq\R^m$, we define the set $\A:= \{u:\Z^n\to X\}$ and adopt the notation $u^i=u(i)$. Interpreting the function $u$ as a $\Z^{n}$-labelled sequence we also write $\{u^{j}\}_{j\in \Z^n}$ or $\{u^{j}\}_{j}$, so that $\{u^{i+j}\}_{j\in \Z^n}$ stands for the translation of $u$ by $i$.

We recall that a measurable subset $A$ of $\RR^n$ is of finite perimeter if the distributional derivative of its characteristic function $\chi_A$ is a bounded measure. In this case, we will simply denote by $\partial A$ the reduced boundary of $A$ and by $\nu\in {\cal S}^{n-1}$ its inner normal, defined by
$$
D\chi_A(B)= \int_{B\cap \partial A} \nu\, d\H^{n-1}
$$
for all Borel set $B$. We will use the fundamental compactness property of sets of finite perimeter:
if $\{A_k\}$ is a sequence of equibounded perimeter; i.e, $\H^{n-1}(\partial A_k)\le C<+\infty$, then
there exists a subsequence $\{A_{k_j}\}$ such that $\chi_{A_{k_j}}$ converge in $L^1_{\rm loc} (\RR^n)$.

%


%



\section{Analysis of discrete energies}\label{Sect:main}
Given $\Omega\subseteq\R^n$ an open set, for all $\e>0$ we introduce the family of functionals $F_\e:{\mathcal A}\to[0,+\infty)$ defined as:
\begin{eqnarray}\label{def:en_latt}
F_\e(u,\Omega):=\sum_{i\in\Z^n\cap \frac{1}{\e}\Omega
}\e^{n-1}\phi(\{u^{i+j}\}_{j\in \Z^n}),
\end{eqnarray}
where $\phi:(X)^{\Z^n}\to[0,L]$ is the potential energy density of the system. In the case $\Omega=\R^n$ we drop the dependence of the energy on the set and simply write $F_\e(u)$ in place of $F_\e(u, \R^n)$.

Note that $F_\e$ is simply a scaling of the energy $F$ defined in \eqref{ef}. More precisely, if we set 
\begin{eqnarray}\label{ef-omega}
F(u,\Omega):=\sum_{i\in\Z^n\cap \Omega}\phi(\{u^{i+j}\}_{j\in \Z^n}),
\end{eqnarray}
then we have 
$$
F_\e(u,\Omega)= \e^{n-1} F\Bigl(u,{1\over\e}\Omega\Bigr).
$$

\subsection{Assumptions on the energy density}\label{Assumption-1}

Given $h\in\N$, we say that $v\in{\mathcal A}$ is $h${\em -periodic} if for all $i\in\{1,2,\dots,n\}$ 
$$v^{j+he_i}=v^j$$ for all $j\in\Z^n$. We assume that there exist $K$ $h$-periodic functions $v_1,v_2,\dots,v_K\in{\mathcal A}$ such that
\begin{itemize}
\item[(H1)] (existence of ground states) we have  $\phi(\{z^{j+i}\}_{j\in\Z^n})=0$  for all $i\in\Z^{n}$ if and only if there exists $l\in\{1,2,\dots,K\}$ such that 
$z^j=v_l^j$  for all $j\in\Z^{n}$;
\item[(H2)] (coerciveness) there exist $ M, M'\in\N$ with $M'\geq M\geq 2$ and $C_M>0$ such that if $u\in\A$ and $i\in\Z^{n}$ are such that for all $l\in\{1,2,\dots, K\}$ there exists $j\in i+MhQ$ such that $u^{j}\neq v_{l}^{j}$ then there exists $i_{M}\in i+M'hQ$ such that 
$$
\phi(\{u^{i_{M}+j}\}_{j\in\Z^n})\geq C_{M};
$$

\item[(H3)] (mild non-locality) given $z,w\in{\mathcal A}$ and $m\in\N$ such that $z^j=w^j$ for all $j\in m Q\cap\Z^n$, then 
$$
|\phi(\{w^j\}_{j\in\Z^n})-\phi(\{z^j\}_{j\in\Z^n})|\le c_m
$$
where the constants $c_m$ are such that $$\sum_{m\in\N}c_mm^{n-1}<+\infty.$$
\end{itemize}

\begin{remark}[general lattices]\rm  For simplicity of notation we parameterize our functions on the lattice $\ZZ^n$, but the result is independent from the choice of the lattice, with another lattice $\cal L$ replacing $\ZZ^n$ in (H1)--(H3). If $\cal L$ is invariant under translations by its elements then the proof remains unchanged since we may still use the notation $\{u^{i+j}\}_j$ as defining a function on $\cal L$.
Otherwise, one has to slightly extend the proof to inhomogeneous energy densities (see Section \ref{inho}).
\end{remark}

\begin{remark}[comments on the hypotheses]\rm  
Hypothesis (H1) and the positiveness of $\phi$ entail that the energy density be chosen and normalized in such a way that it is not minimized pointwise on a function $u$ which is not a periodic minimizer. Note that the same $F$ may be rewritten in terms of many $\phi$. 
For example, a system of nearest and next-to-nearest neighbor interactions in the square lattice for spin systems $X=\{\pm1\}$ can be described by the energy density
\begin{eqnarray*}
&&\hskip-.7cm\phi(\{u^j\}_j)= \sigma_n \Bigl(u^{(1,0)}u^{(0,0)}+  u^{(0,1)}u^{(0,0)}+ u^{(-1,0)}u^{(0,0)}+  u^{(0,-1)}u^{(0,0)}\Bigl)\\
&& \qquad\ \ \ 
 +\sigma_{nn}\Bigl(u^{(1,1)}u^{(0,0)}+  u^{(-1,1)}u^{(0,0)}+ u^{(-1,1)}u^{(0,0)}+  u^{(-1,-1)}u^{(0,0)}+4\Bigl) 
\end{eqnarray*}
with $\sigma_n$, $\sigma_{nn}$ positive.
In this case $\phi(\{u^j\}_j)= -4\sigma_n<0$ if $u^{(0,0)}=1$ and $u^i=-1$ if $i\neq(0,0)$,
so that the positiveness assumption is not satisfied.
We can anyhow regroup the interactions in such a way that the energy $F$ be described by a new positive $\phi$ satisfying (H1)--(H3) (see Example \ref{ABC_ex} and \cite{ABC}).

Hypothesis (H2) guarantees that if a function differs from a ground state in a suitably large cell, then we have some positive contribution within a fixed distance of that cell. Note that we have 
to take into account the possibility of a larger cell by the non-local nature of discrete interaction.
As an example we may take one-dimensional spins on $\ZZ$ with the energy density
$$
\phi(\{u^j\}_j)= (u^3-u^0)^2+(u^5-u^3)^2.
$$
Since $3$ and $5$ are coprime, the uniform states are the only ground states, so that we may take $h=1$ and $M=2$. The condition on the function $u$ in (H2) is then that it satisfies $u^0\neq u^1$. If $\phi(\{u^j\}_j)=\phi(\{u^{j+1}\}_j)=0$ then $u^3=u^0\neq u^1=u^4=u^6$ so that 
$\phi(\{u^{j+3}\}_j)\ge 4$, and 
we may take $M'=3$  and $C_M=4$.

Hypothesis (H3) implies that if two functions are equal on a regular set then their difference of energy on that set can be estimated by a sum of terms decaying with the distance from the boundary of that set, which allows to prove a locality property. 
The decay condition on $c_m$ can be interpreted as the requirement that
a single non-minimal inclusion give a finite contribution. This condition can be relaxed 
when $\Omega$ is bounded since in that case we can assume that the $\H^{n-1}$ measure 
of level sets 
of the distance from a boundary can be estimated by the measure of the boundary itself.
In that case, the decay condition is
\begin{equation}
\sum_{m\in\N}c_m<+\infty.
\end{equation}
This corresponds to the decay condition for ferromagnetic spin systems in \cite{AlGe}.

\end{remark}

\begin{remark}\label{stima}\rm
By (H1) and (H3) a function $z\in\A$ such that $z^j=v_l^j$ for all $j \in (i+m Q)\cap\Z^n$ for some $m\in\N$ satisfies
\begin{equation}\label{r:mild_nonloc}
\phi(\{z^{i+j}\}_{j\in\Z^n})\le c_m
\end{equation}
with $c_m$ as in (H3).
\end{remark}

\begin{remark}\rm
Since the energy is invariant under translation, by (H1) the functions $w_l$ defined as $w_l^j=v_l^{j+j'}$ for some $j'\in\Z^n$ are ground states of the system; i.e., 
every pattern gives a family of modulated phases indexed by the possible translations.
By (H2) we have that, if $v_l^j=v_{l'}^{j+j'}$ for some $j'\in\Z^n$, then $l=l'$. \end{remark}

\subsection{From functions to sets: choice of the convergence}\label{convo}
In what follows to each function belonging to $\A$ we associate a $K$-ple of sets and define a notion of convergence accordingly. 

Given $u\in\A$ and $l\in\{1,2,\dots,K\}$, we define 
\begin{eqnarray}\label{def:reticolo di fase}
I_{l}(u):=\{ j\in\Z^n:\ u^{ i}=v_l^{ i},\ \hbox{ for all } i\in (jh+hQ)\cap\Z^n\}
\end{eqnarray} 
and define the {\sl phase} $l$ as the set 
$$
A_{\e,l}(u):=\bigcup\limits_{j\in I_{l}(u)}\e(jh+hQ).
$$

\begin{definition}[discrete-to-continuum convergence]\label{convergence}
Let $u_\e\in\A$ be a sequence of discrete functions and let ${\mathbb A}=(A_1,A_2,\dots,A_K)\subseteq(\R^n)^K$. 
We say that $u_\e$ {\em  converges to }$\mathbb A$ and we write $u_\e\to {\mathbb A}$ if  $|(A_{\e,l}(u_{\e})\Delta A_{l})\cap RQ|\to 0$
for all $l\in\{1,2,\dots,K\}$ and for all $R>0$.
\end{definition}

\begin{remark}\rm
To any K-ple of sets $\mathbb{A}=(A_1,A_2,\dots,A_K)\subseteq(\R^n)^K$ we associate the function $p({\mathbb A}):\R^n\to\{0,1,2,\dots,K\}$ defined as
\begin{eqnarray*}
p_{\mathbb A}(x):=\sum_{l=1}^Kl\chi_{A_l}(x).
\end{eqnarray*}
Note that whenever $\mathbb{A}=(A_1,A_2,\dots, A_K)$ is a partition of $\R^n$ into sets of finite perimeter, then 
the convergence $u_{\e}\to \mathbb{A}$ is equivalent to the convergence of the functions 
\begin{equation}\label{pconvergence}
p(u_{\e}):=p_{{\mathbb A_\e}(u_\e)},
\end{equation}
where ${\mathbb A_\e}(u_\e)=(A_{\e,1},A_{\e,2},\dots,A_{\e,K})$, to $p_{\mathbb A}$ in $L^1_{\rm loc}(\R^n)$. 
\end{remark}

\subsection{Discrete-to-continuum analysis}
In this section we state our main result regarding the asymptotic behavior of the energies in \eqref{def:en_latt} as $\e\to 0$. 


In what follows we prove that, up to subsequences, as $\e\to 0$ the phases of a discrete systems with equibounded energy form a partition of $\R^n$ into sets of finite perimeter.
\begin{theorem}[compactness]\label{teo:comp}
Let $\phi$ satisfy hypothesis {\rm (H2)}. Let $u_\e\in\A$ be such that $\sup_\e F_\e(u_\e)<+\infty$. Then, up to subsequences, $u_\e$ converges to ${\mathbb A}:=(A_1,A_2,\dots,A_K)$  where, for all $l\in\{1,2,\dots,K\}$, $A_l$ is a set of 
finite perimeter in $\R^n$ and the sets $A_1,A_2,\dots,A_K$ form a partition of $\R^n$.
\end{theorem}
\begin{proof}
Let $I_{\e,l}(u_{\e})$ be defined as in \eqref{def:reticolo di fase} and let 
$$
N_l(u_\e):=\{i\in I_{\e,l}(u_\e):\ \exists j\not\in I_{\e,l}(u_\e),\  \|i-j\|_{\infty}=M\}.
$$ 
By $(H2)$ we have that for all $i\in N_l(u_\e)$ 
$$
F_\e(u_\e, ih+M'Q_h)\geq C_{M}\e^{n-1}.
$$
Hence the following estimate holds:
$$
\#N_l(u_\e)\e^{n-1}\leq \frac{1}{C_M} \sum_{i\in N_l(u_\e)}F_\e(u_\e, ih+M'Q_h)\leq \frac{(M')^{n}}{C_{M}}F_{\e}(u_{\e}).
$$
Therefore, for all $l\in\{1,2,\dots,K\}$, it holds that
\begin{eqnarray}\label{Tcomp:stima_2}
\H^{n-1}(\partial (A_{\e,l}(u_\e)))&\leq& 2n h^{n-1}\e^{n-1}\#N_l(u_\e)\nonumber\\
&\leq& \frac{2n h^{n-1}(M')^{n}}{C_{M}} F_{\e}(u_{\e})<\infty.
\end{eqnarray}
Estimate \eqref{Tcomp:stima_2} implies that each family $\{A_{\e,l}(u_\e)\}_\e$
has equibounded perimeter, so that its characteristic functions are precompact in $L^1_{\rm loc}$. Hence, we obtain the convergence of a subsequence of $u_\e$ to some ${\mathbb A}=(A_1,A_2\dots, A_K)$. 

We now prove that the sets $A_1,A_2,\dots,A_K$ are a partition of $\R^n$. Let us set $A_\e=\bigcup_{l=1}^KA_{\e,l}(u_\e)$ and $I_\e=\Z^{n\setminus }\bigcup_{l=1}^KI_{\e,l}(u_\e)$. 
For all $i\in I_{\e}$ let $i_{M}(i)$ be given by $(H2)$. We have that 
\begin{eqnarray}\label{Tcomp:stima_1}\nonumber
|\R^n\setminus V_\e|&\leq& \e^n\#(I_\e)\leq \frac{\e^{n}}{C_{M}}\sum_{i\in I_{\e}}\phi(\{u_{\e}^{i_{M}(i)+j}\}_{j\in \Z^n})\\&\leq& \frac{\e}{(M')^{n}C_{M}}F_{\e}(u_{\e})\leq \e C,
\end{eqnarray}
where we have taken into account that $i_{M}(i)$ belongs to $(M')^{n}$ cubes of side length $M$. The thesis follows letting $\e$ tend to $0$.
\end{proof}

In order to state our main theorem we have to give a meaning to boundary conditions for our energies. To that end, usually functions are set as fixed outside the domain. In our case we further specify their value in a small zone in the interior, which will turn out useful to cope with the non-locality of the energy densities. For all open sets $\Omega$, $u_0\in\A$ and $\eta>0$ (the thickness of the `safe zone') we set
\begin{equation}\label{boundary}
\Omega_\eta=\{ x\in\Omega: {\rm dist}(x,\R^n\setminus \Omega)>\eta\}
\end{equation}
and 
\begin{equation}\label{conbo}
{\mathcal B}^{u_0}_\eta(\Omega)=\{ u\in\A: u=u_0 \hbox{ outside } \Omega_\eta\}.
\end{equation}

We will use boundary conditions when $\Omega$ is a cube and $u_0$ is a function taking only two phases, defined as follows. For $\nu\in{\mathcal S}^{n-1}$ we set
\begin{equation}
\Pi_{\nu}^+= \bigcup_j\Bigl\{ hj+hQ : j\in\Z^n, (j,\nu)\ge 0\Bigr\}\cap \Z^n.
\end{equation}
For $\nu\in{\mathcal S}^{n-1}$ and $l,l'\in\{1,2,\dots,K\}$ we then define $u_{l,l'\!\!,\nu}\in\A$ as
\begin{equation}\label{ull}
u_{l,l'\!\!,\nu}^i=\begin{cases}
v^i_l &\hbox{ if } i\in \Pi_{\nu}^+\cr
v^i_{l'} &\hbox{ otherwise.}
\end{cases}
\end{equation}
Note that the sequence $u_\e=u_{l,l'\!\!,\nu}$ converges to $(A_1,\dots, A_K)$, where
$A_l=\{x:  (x,\nu)> 0\}$, $A_{l'}=\{x:  (x,\nu)< 0\}$, and $A_j=\emptyset$ for $j\not\in\{l,l'\}$.

We finally introduce, for $\e,\d, T>0$, $l,l'\in\{1,2,\dots,K\}$ and $\nu\in{\mathcal S}^{n-1}$ the class of discrete functions
\begin{eqnarray}
{\mathcal B}^{l,l'\!\!,\nu}_{\d}(TQ_\nu):= {\mathcal B}^{u_{l,l'\!\!,\nu}}_{\delta T}(TQ_\nu)
\end{eqnarray}
according to the notation in \eqref{conbo}.

\begin{theorem}[$\Gamma$-convergence]\label{teo:Gamma}
Let the energy functionals $F_\e$ satisfy hypotheses {\rm (H1)--(H3)} and $\Omega$ be a Lipschitz set. Then, there exists the $\Gamma$-limit of $F_\e(\cdot,\Omega)$ as $\e\to 0$ with respect to the convergence in Definition {\rm\ref{convergence}},  and we have 
\begin{eqnarray}\label{teo:Gammalim-for}
\Gamma\hbox{-}\lim_{\e\to 0}F_\e(u,\Omega)={\cal F}({\mathbb A},\Omega):=\sum_{l,l'=1}^K\int_{\Omega\cap \partial A_l\cap\partial A_{l'}}\varphi(l,l',\nu_{l,l'})\,d\H^{n-1},
\end{eqnarray}
for all ${\mathbb A}:=\{A_1,A_2,\dots,A_K\}$ partitions of $\R^n$ into sets of finite perimeter,
where $\varphi:\{1,\dots,K\}^2\times {\mathcal S}^{n-1}\to [0,+\infty)$ satisfies
\begin{eqnarray}\label{teo:Gamma-lim}\nonumber
\varphi(l,l',\nu)
&=&\lim_{\d\to 0^+}\liminf_{T\to +\infty}\frac{1}{T^{n-1}}\inf\{F(u,TQ_{\nu}),\ u\in{\mathcal B}^{l,l'\!\!,\nu}_{\d}(TQ_\nu)\}
\\ 
&=&\lim_{\d\to 0^+}\limsup_{T\to +\infty}\frac{1}{T^{n-1}}\inf\{F(u,TQ_{\nu}),\ u\in{\mathcal B}^{l,l'\!\!,\nu}_{\d}(TQ_\nu)\}.\qquad\qquad
\end{eqnarray}
\end{theorem}
\begin{remark}{\rm Note that the existence of the limit in $\delta$ in formula \eqref{teo:Gamma-lim} is a consequence of the inclusion ${\mathcal B}^{u_0}_{\eta}(\Omega)\subseteq{\mathcal B}^{u_0}_{\eta'}(\Omega)$ for $\eta<\eta'$ which implies that this limit is actually an infimum.}
\end{remark}

The proof of the result will be subdivided in two steps, first proving a compactness and integral representation result, and then identifying the limit energy density through homogenization formulas giving the characterization (\ref{teo:Gamma-lim}). 

\subsection{Simplified assumptions on the energy density}\label{simply}
The proof of Theorem \ref{teo:Gamma} will be performed under a simplified set of hypotheses detailed below. We may indeed suppose in (H1)--(H3) that $h=1$, $2M= M'$ and that $v_{l}$ are constant. Namely, 
\begin{itemize}
\item[(H1)] (existence of constant ground states) we have $\phi(\{z^{j+i}\}_{j\in\Z^n})=0$ for all $i\in\Z^{n}$ if and only if  $l\in\{1,2,\dots,K\}$ exists such that 
$z^j=v_l$ for all $j\in\Z^{n}$;
\item[(H2)] (coerciveness on nearest-neighbors) there exists $C>0$ such that 
if $u_0\neq u_i$ for some $i\in\Z^n$ with $\|i\|_\infty=1$, then there exists $j$ with $\|j\|_\infty\le 2$ such that 
$$
\phi(\{u^{j+k}\}_{k\in\Z^n})\geq C;
$$
\item[(H3)] (mild non-locality) given $z,w\in{\mathcal A}$ and $m\in\N$ such that $z^j=w^j$ for all $j\in m Q\cap\Z^n$, then 
$$
|\phi(\{w^j\}_{j\in\Z^n})-\phi(\{z^j\}_{j\in\Z^n})|\le c_m
$$
where the constants $c_m$ are such that $\sum_{m\in\N}c_mm^{n-1}<+\infty$.
\end{itemize}
We show that this choice does not affect the generality upon the introduction of a possibly much larger set of parameters. 

We suppose that $\phi$ satisfies assumptions (H1)--(H3) in Section \ref{Assumption-1}. It is not restrictive to suppose that $M\in 2h\Z$  and that $M'\leq 2M$. We will construct an equivalent energy by constructing a suitable energy density $\tilde\phi$. 

Let $\tilde X=(X)^{MQ}$ and define the bijection $\Psi:(\Z^{n})^{\tilde X}\to (\Z^{n})^{X}$ as follows. Given $\tilde u:\Z^{n}\to \tilde X$ 
\begin{equation}
(\Psi(\tilde u))^{i}=(\tilde u^{\tilde i})^{i-\tilde i M},\quad \hbox{ for all }i\in\tilde iM+MQ. 
\end{equation}
We then define $\tilde \phi:(\Z^{n})^{\tilde X}\to [0,L]$ as
\begin{equation}
\tilde \phi(\{\tilde u^{\tilde j}\}_{\tilde j})=\frac{1}{M^{n}}\sum_{i\in MQ}\phi(\{(\Psi(\tilde u))^{i+j}\}_{j}).
\end{equation}
(H1) is satisfied with $\tilde v_{l}=\Psi^{-1}(v_{l})$ in place of $v_{l}$, and (H3) with $c_{m}$ replaced by $\tilde c_{m}=c_{(m-1)M}$ (taking $c_{0}=2L$). As for (H2), it holds with $C=C_{M}M^{-n}$. Indeed, if $\tilde u^0\not\in\{\tilde v_1,\ldots, \tilde v_K\}$ then we can use the corresponding hypothesis in Section \ref{Assumption-1} with $i=0$,
while if $\tilde u^0\in \{\tilde v_1,\ldots, \tilde v_K\}$ then we use the same hypothesis with $i={M\over 2}e_1$,
noting that the condition $\tilde u^0\neq  \tilde u^1$ implies that $u$ does not coincide with a ground state
in ${M\over 2}e_1+MQ$.

We note that 

(i) setting 
$$
\tilde F_{\e}(\tilde u,A)=\sum_{\tilde j\in(\frac{1}{\e M}A\cap Z^{n})}\e^{n-1}\tilde\phi(\{\tilde u^{\tilde j}\}_{\tilde j})
$$
we have that $F_{\e}(u,A)=\tilde F_{\e}(\Psi^{-1}(u),A)$;

(ii) we may define a convergence $\tilde u_{\e}\to\mathbb A$ as in Section \ref{convo}. In this case the sets $I_{l}(\tilde u)$ are simply defined as
$I_{l}(\tilde u)=\{i\in\Z^{n}:\, \tilde u^{i}=\tilde v_{l}\}$ and the definition of the sets $\tilde A_{\e,l}$ analog to the sets $A_{\e,l}$ must take into account the scaling factor $M$; namely, 
$$
\tilde A_{\e,l}(\tilde u)=\bigcup_{i\in I_{l}} \e(iM+MQ).
$$

By Theorem \ref{teo:comp} with $\tilde \phi$ in place of $\phi$, this convergence is compact and if  $u_{\e}\to\mathbb A$ then $\Psi^{-1}(u_{\e})$ converges to the same $\mathbb A$. Indeed by construction $\tilde A_{\e,l}(\Psi^{-1}(u_{\e}))\subseteq A_{\e,l}(u_{\e})$ so that the corresponding inclusions hold in the limit. The conclusion follows remarking that we must have equality since the limiting sets are partitions. 

Finally we note that, by (i) and (ii), the $\Gamma$-convergence of $F_{\e}$ with respect to the convergence $u_{\e}\to \mathbb A$ is equivalent to  
the $\Gamma$-convergence of $\tilde F_{\e}$ with respect to the convergence $\tilde u_{\e}\to \mathbb A$.

\begin{remark}\rm
Note that, if $\tilde X$ is identified with a subset of an Euclidean space, then the convergence $\tilde u_{\e}\to \mathbb A$ is equivalent to the $L^{1}_{\rm loc}$-convergence of their piecewise-constant interpolations to $\sum_{l=1}^{K}v_{l}\chi_{A_{l}}$.  This shows that the $\Gamma$-limit can be set in the framework of separable metric spaces, and that it enjoys lower-semicontinuity properties with respect to the $L^1_{\rm loc}$-convergence of the elements of partitions.
\end{remark}

\section{Proof of the theorem}
In this section we will work in the simplified set of hypotheses of Section \ref{simply}, taking into account the observations therein. We will follow an indirect argument by first proving an integral representation result and then identifying the energy density. The necessity of this type of proof derives from the lack of a strong density result of polyhedral (or smooth) partitions in all Caccioppoli partitions, which makes the construction of recovery sequences particularly laborious.

\subsection{Compactness and integral representation}

We will follow the localization method for $\Gamma$-convergence \cite{GCB} by studying the properties of $F_\e(u,\Omega)$ in dependence of the set $\Omega$. The following lemma will be used in particular to prove that the $\Gamma$-limit $F_0({\mathbb A},\Omega)$ of (a subsequence of) these energies is the restriction of a measure for fixed $\mathbb A$, and to deal with boundary-value problems. 

\begin{lemma}\label{stimaf}
Let $U'\subset\subset U\subset\subset V$ be bounded open sets in $\R^n$ with $U'$ Lipschitz and let $u_{U,\e}$ and $u_{V,\e}$ be sequences such that  
\begin{equation}\label{lemma:bound}
\sup_\e(F_\e(u_{U,\e})+F_\e(u_{V,\e}))\leq C<+\infty.
\end{equation}
Then for all $R\in \N$ there exists $v_\e\in\A$ such that\\

{\rm(i)} $v_{\e}=u_{{U,\e}}$ on $\Z^{n}\cap \frac{1}{\e}U'$,  \\

{\rm(ii)} $v_{\e}=u_{{V,\e}}$ on $\Z^{n}\setminus \frac{1}{\e}U$,\\

{\rm(iii)} for all $V'\subseteq V$ 

\begin{eqnarray}\label{stimaf-formula}\nonumber
F_\e(v_{\e}, V')&\le& F_\e(u_{{U,\e}}, V'\cap U)+F_\e(u_{{V,\e}}, V'\setminus \overline {U'})\\
&&+C_{UV}\e R^{n+1}\Bigl(F_\e(u_{{U,\e}}, U)+F_\e(u_{{V,\e}}, V)\Bigr) \nonumber \\
&&+C_{UV}\Bigl(\sum_{\alpha\geq R}c_{\alpha}
+R^{n+1}\|p(u_{U,\e})-p(u_{V,\e})\|_{L^{1}(U\setminus U')}\Bigr),\qquad
\end{eqnarray}
where $C_{UV}$ denotes a constant depending only on $U$ and $V$, $p(u)$ is defined in \eqref{pconvergence} and $c_{\alpha}$ is the same as in 
$\rm (H3)$. Furthermore if $u_{U,\e}$ and $u_{V,\e}$ both converge to $\mathbb A$, then $v_{\e}$ converges to $\mathbb A$. 
\end{lemma}
\begin{proof}
Let $\mu_{\e}^{U},\ \mu_{\e}^{V}\in{\mathcal M}^{+}(\R^{N})$ be defined as
\begin{equation*}\label{measures}
\mu_{\e}^{U}=\sum_{i\in\Z^{n}}\e^{n-1}\phi(\{u^{i+j}_{U,\e}\}_{j})\delta_{\e i}, \qquad \mu_{\e}^{V}=\sum_{i\in\Z^{n}}\e^{n-1}\phi(\{u^{i+j}_{V,\e}\}_{j})\delta_{\e i}.
\end{equation*} 
In terms of $\mu_{\e}^{U}$ and $\mu_{\e}^{V}$ assumption \eqref{lemma:bound} reads
\begin{equation*}
\sup_{\e}(\mu_{\e}^{U}(\R^{n})+\mu_{\e}^{V}(\R^{n}))\leq C<+\infty,
\end{equation*} 
As a result there exist two positive finite measures $\mu^{U},\, \mu^{V}\in{\mathcal M}^{+}(\R^{n})$ such that, up to subsequences,
\begin{equation}\label{lemma:comp-meas}
\mu_{\e}^{U}\rightharpoonup \mu^{U} \hbox{ and } \mu_{\e}^{V}\rightharpoonup \mu^{V}.
\end{equation}
Moreover, by \eqref{lemma:bound} and Theorem \ref{teo:comp}, we also have that there exists ${\mathbb A}_{U}$ and ${\mathbb A}_{V}$ partitions of $\R^{n}$ such that, up to subsequences, 
\begin{equation}\label{lemma:comp-funct}
u_{U,\e}\to{\mathbb A}_{U}\hbox{ and } u_{V,\e}\to {\mathbb A}_{V}.
\end{equation}
By \eqref{lemma:comp-meas} and \eqref{lemma:comp-funct} there exists $c\in(0,{\rm dist}(U\setminus U'))$ such that, setting 
$$
A_{c}:=\{x\in \R^{n}:\, {\rm dist}(U',x)=c\},
$$  
it holds
\begin{eqnarray}\label{lemma:est-small}
\mu_{\e}^{U}(A_{c}+B_{8\sqrt{n}R\e})+\mu_{\e}^{V}(A_{c}+B_{8\sqrt{n}R\e})&\leq& \e R C_{UV}(F_\e(u_{U,\e})+F_\e(u_{V,\e}))\\
\|p(u_{U,\e})-p(u_{V,\e})\|_{L^{1}(A_{c}+B_{8\sqrt{n}R\e})}&\leq& \e R C_{UV}\|p(u_{U,\e})-p(u_{V,\e})\|_{L^{1}(U\setminus U')}.\nonumber
\end{eqnarray}
We set 
\begin{equation}
J_{R,\e}=\{j\in\Z^{n}:\, \e(Rj+RQ)\cap A_{c}\neq\emptyset\}
\end{equation}
and note that 
\begin{equation}\label{lemma:sti-J}
\# J_{R,\e}\leq C_{UV}\e^{1-n}R^{1-n}.
\end{equation} 
We define
\begin{equation}
v_{\e}^{i}=\begin{cases}
u_{V,\e}^{i}& \text{ if } {\rm dist}(U',\e i)>c\, \text{ or if } i\in Rj+RQ\, \text{ for some }j\in J_{R,\e}\\
u_{U,\e}^{i}& \text{ otherwise,}\\
\end{cases}
\end{equation}
and we note that (i) and (ii) hold true.
Setting
\begin{eqnarray*}
J_{R,\e}^{U}&=&\{j\in J_{R,\e}:\, u_{U,\e}\not\equiv v_{l}\, \hbox{on }Rj+4RQ \hbox{ for any } l\in\{1,2,\dots,K\}\},\\
J_{R,\e}^{V}&=&\{j\in J_{R,\e}:\, u_{V,\e}\not\equiv v_{l}\, \hbox{on }Rj+4RQ \hbox{ for any } l\in\{1,2,\dots,K\}\},\\
H_{R,\e}&=&\{j\in J_{R,\e}\setminus(J^{U}_{R,\e}\cup J^{V}_{R,\e}):\, u_{U,\e}\not\equiv u_{V,\e} \hbox{ on }Rj+4RQ\},
\end{eqnarray*}
we note that 
\begin{eqnarray}
\nonumber\bigcup_{j\in J_{R,\e}^{U}}\e(Rj+4RQ)\subset A_{c}+B_{8\sqrt{n}R\e},\\ 
\nonumber\bigcup_{j\in J_{R,\e}^{V}}\e(Rj+4RQ)\subset A_{c}+B_{8\sqrt{n}R\e},\\
\nonumber\bigcup_{j\in H_{R,\e}}\e(Rj+4RQ)\subset A_{c}+B_{8\sqrt{n}R\e}.
\end{eqnarray}
From the first two inclusions, by $(H2)$ and \eqref{lemma:est-small} we get
\begin{equation}\label{lemma:card-bound1}
\e^{n-1}(\# J_{R,\e}^{U}+\# J_{R,\e}^{V})\leq \e RC_{UV}(F_\e(u_{U,\e})+F_\e(u_{V,\e})),
\end{equation}
while from the third inclusion using \eqref{lemma:est-small} we obtain
\begin{equation}\label{lemma:card-bound2}
\# H_{R,\e}\e^{n}\leq \e R C_{UV}\|p(u_{U,\e})-p(u_{V,\e})\|_{L^{1}(U\setminus U')}.
\end{equation}
Setting $D_{c}=\{x:\, {\rm dist}(U',x)\leq c\}$ we define 
\begin{eqnarray}\nonumber
U_{R,\e}=\bigcup\{\e(Rj+RQ):\, \e(Rj+3RQ)\cap A_{c}=\emptyset,\, \e(Rj+3RQ)\subset D_{c}\},\\
\nonumber V_{R,\e}=\bigcup\{\e(Rj+RQ):\, \e(Rj+3RQ)\cap A_{c}=\emptyset,\, \e(Rj+3RQ)\subset \R^{n}\setminus D_{c}\}.
\end{eqnarray}
For all $V'\subset V$ we may write 
\begin{eqnarray}\label{lemma:est}\nonumber
F_{\e}(v_{\e}, V')&\leq& F_{\e}(v_{\e}, U_{R,\e}\cap V')+F_{\e}(v_{\e}, V_{R,\e}\cap V')
\\
&&+F_{\e}(v_{\e}, V'\setminus (U_{R,\e}\cup V_{R,\e})).
\end{eqnarray}
Note that if $j\in U_{R,\e}/\e$ then $v_{\e}^{i}\equiv u^{i}_{U,\e}$ for all $i\in j+\alpha_{\e}(j)Q$ with $\e \alpha_{\e}(j)-{\rm dist}(\e j, A_{c})=o(1)$, so that $\alpha_{\e}(j)$ diverges as $\e\to 0$. Combining that with $(H3)$ and the positivity of the energy, we have
\begin{eqnarray}\label{lemma:est1}
\nonumber F_{\e}(v_{\e}, U_{R,\e})&\leq& F_{\e}(u_{U,\e}, U_{R,\e})+\sum_{\alpha\geq\alpha_{\e}(j)}\e^{n-1}c_{\alpha}\#\{j\in U_{R,\e}:\, \alpha_{\e}(j)=\alpha\}\\
&\leq&F_{\e}(u_{U,\e},U)+C_{UV}\sum_{\alpha\geq R}c_{\alpha}
+o(1).
\end{eqnarray}
Note that here we use the hypothesis that $U'$ be Lipschitz.
With the same argument, one has
\begin{equation}\label{lemma:est2}
F_{\e}(v_{\e}, V_{R,\e}\cap V')\leq F_{\e}(u_{V,\e},V\setminus \overline U')+C_{UV}\sum_{\alpha\geq R}c_{\alpha}
+o(1).
\end{equation}
Setting
\begin{eqnarray*}
Z_{R,\e}&=&\bigcup\{\e(Rj+4RQ):\, j\in J_{R,\e}\},\\
W_{R,\e}&=&\{j\in\Z^{n}:\, \e(Rj+RQ)\subseteq Z_{R,\e}\},\\
\overline H_{R,\e}&=&\{j\in Z^{n}:\,Rj+RQ\subset Rz+4RQ  \hbox{ for some } z\in H_{R,\e}\},\\
\overline J_{R,\e}&=&\{j\in Z^{n}:\,   Rj+RQ\subset Rz+4RQ \hbox{ for some }
z\in J_{R,\e}^{U}\cup J_{R,\e}^{V}\},
\end{eqnarray*}
using $(H3)$, \eqref{lemma:card-bound1}, \eqref{lemma:card-bound2} and the fact that, by \eqref{lemma:sti-J}, $\#W_{R,\e}\leq C_{UV}\e^{1-n}R^{1-n}$, we may write 
\begin{eqnarray}\label{lemma:est3}
\nonumber F_{\e}(v_{\e}, V'\setminus (U_{R,\e}\cup V_{R,\e}))&\leq& \nonumber F_{\e}(v_{\e}, Z_{R,\e})
\\ \nonumber &\leq&\sum_{j\in \overline J_{R,\e}}F_{\e}(v_{\e}, \e j+\e RQ)\\&&\nonumber +\sum_{j\in \overline H_{R,\e}}F_{\e}(v_{\e}, \e j+\e RQ)\\
\nonumber &&+\sum_{j\in W_{R,\e}\setminus(\overline J_{R,\e}\cup \overline H_{R,\e})}F_{\e}(v_{\e}, \e j+\e RQ)\\
\nonumber &\leq&4^{n}LR^{n}(\# J_{R,\e}^{U}+\# J_{R,\e}^{V})\e^{n-1}\\&&\nonumber
+4^{n}LR^{n}(\# H_{R,\e})\e^{n-1}
+R^{n}\sum_{j\in W_{R,\e}}c_{R}\e^{n-1}\\ \nonumber
&\leq&\e R^{n+1}C_{UV}(F_\e(u_{U,\e})+F_\e(u_{V,\e}))\\&&\nonumber+R^{n+1}C_{UV}\|p(u_{U,\e})-p(u_{V,\e})\|_{L^{1}(U\setminus U')}\\&&+C_{UV}RC_{R}.
\end{eqnarray}
Eventually, we get \eqref{stimaf-formula} by gathering inequalities \eqref{lemma:est1}, \eqref{lemma:est2} and \eqref{lemma:est3} together in \eqref{lemma:est}.
\end{proof}

\begin{remark}\label{appro-polse}\rm
For all $\mathbb A$ there exists a partition composed of polyhedral sets $\mathbb A^\e$
such that $|(A_l^\e\triangle A_l)\cap RQ|\to 0$ for all $R$, 
$$
\limsup_\e\sum_{l=1}^K\H^{n-1} (\partial A^\e_l)\le KC\sum_{l=1}^K\H^{n-1} (\partial A_l)
$$
and $(\partial A_l^\e\cap \partial A_{l'}^\e)\cap (\partial A_m^\e\cap \partial A_{m'}^\e)=\emptyset$
if $\{l,l'\}\neq\{m,m'\}$.

Indeed, upon identifying a partition $\mathbb A$ with a function $u\in BV_{\rm loc}(\RR^n)$ by setting $u(x)=l$ if $x\in A_l$ we can use a classical argument by computing a mollified $u_\rho$ and using the coarea formula to find a partition $\mathbb A^\rho $ composed of $C^\infty$ sets converging to $\mathbb A$ as $\rho\to 0$ and with 
$$
\sum_{l=1}^K \H^{n-1}(\partial A^\rho_l)\le |Du|(\RR^n)\le K\sum_{l=1}^K \H^{n-1}(\partial A_l).
$$
(see \cite{BBR} for a detailed argument). The desired partitions can be obtained by separately approximating each $A^\rho_l$ by polyhedral sets.
\end{remark}

\begin{proposition}\label{upperbound}
Let $F_{\e}$ satisfy the same assumption of Theorem {\rm\ref{teo:Gamma}} and let $\Omega$ be an open Lipschitz set in $\R^{n}$. Then there exists a constant $C>0$ such that 
\begin{equation}\label{upperbound-statement}
\Gamma\hbox{-}\limsup_{\e}F_{\e}(\mathbb A,\Omega)\leq C\sum_{l=1}^{K}{\mathcal H}^{n-1}(\Omega\cap\partial A_{l}),
\end{equation}
where $\Gamma\hbox{-}\limsup_{\e}F_{\e}(\mathbb A,\Omega)$ denotes the $\Gamma\hbox{-}\limsup$ of the functionals $F_{\e}(u_{\e},\Omega)$ with respect to the convergence of $u_{\e}$ to $\mathbb A$.  
\end{proposition}
\begin{proof}
In the course of the proof we will use the property that $$\mathbb A\mapsto \Gamma\hbox{-}\limsup_{\e}F_{\e}(\mathbb A,\Omega)$$ is a lower semicontinuous function with respect to the $L^{1}$-convergence of sets, so that we are allowed small variations of our target partition (see \cite{GCB} Remark 1.29).

\smallskip
We first consider the case $\Omega=\R^{n}$ and $\mathbb A$ a partition of $\R^{n}$ into polyhedral sets, and preliminarily deal with the case of a partition composed of only two polyhedral sets. We can suppose that $A_{1}=A$, $A_{2}=\R^{n}\setminus A$ and $A_{k}=\emptyset$ for all $k> 2$. 

\def\I{{\mathcal I}}
\def\J{{\mathcal J}}

We decompose $\partial A$ as follows
\begin{equation}
\partial A=\bigcup_{l\in \I}V_{l},
\end{equation}
with $V_{l}$ is closed $n-1$-dimensional polytope and where $\I$ is a finite set of indices and $V_{l}\cap V_{l'}$ is a $n-2$-dimensional polytope. 
We denote by $\nu_{l}$ the exterior normal to $A$ at $V_{l}$. Up to a translation we may suppose that $\frac{1}{\e}\partial A\cap \Z^{n}=\emptyset$; furthermore up to a rotation of $A$ we may also suppose that $(\nu_{l},e_{k})\neq 0$ for all $k\in\{1,2,\dots n\}$ and $l\in\I$, so that 
for all $x\in\R^{n}$ there exists a unique $\overline x\in\Pi_{\nu_{l}}$ such that $\|\overline x-x\|_{\infty}=\min\{\|y-x\|_{\infty},\, y\in \partial A\}$.
Note that, if we define $\xi_{l}$ as the vector such that $(\xi_{l}, e_{k})={\rm sign} (\nu_{l},e_{k})$, then $(x-\overline x)$ is parallel to $\xi_{l}$. 
We define $v_{\e}\in\A$ as
\begin{equation}
v^{i}_{\e}=\begin{cases}	
v_{1}& \hbox{ if }\e i\in A\\
v_{2}& \hbox{ otherwise,}
\end{cases}
\end{equation} 
where $v_{1}$ and $v_{2}$ are as in $(H1)$ and observe that $v_{\e}\to \mathbb A$.
We set
\begin{equation}
\alpha_{\e}(i)=\sup\{\alpha:\, v_{\e} \hbox{ is constant on }(\e i+\e \alpha Q)\},
\end{equation}
$S_{l}=V_{l}+\xi_{l}\R$ and define the set of indices 
\begin{equation}
\J_{l,\e}=\{i\in \Z^{n}\cap \frac{1}{\e}S_{l}:\, \e(i+(\alpha_{\e}(i)+1)Q)\cap V_{l}\neq\emptyset\}.
\end{equation}
We note that 
\begin{equation}
\# \{i\in\J_{l,\e}:\,\alpha_{\e}(i)=\alpha\}\leq \frac{C}{\e^{n-1}}(\H^{n-1}(V_{l})+o_{\e}(1)).
\end{equation}
We now estimate the energy of $v_{\e}$ as
\begin{eqnarray}\label{upperbound}
F_{\e}(v_{\e})=\sum_{l\in\I}\sum_{i\in\J_{l,\e}}\e^{n-1}\phi(\{v_{\e}^{i+j}\}_{j})+\sum_{i\not\in\bigcup_{l\in\I}\J_{l,\e}}\e^{n-1}\phi(\{v_{\e}^{i+j}\}_{j}).
\end{eqnarray}
 \begin{figure}[h!]
\centerline{\includegraphics [width=2.5in]{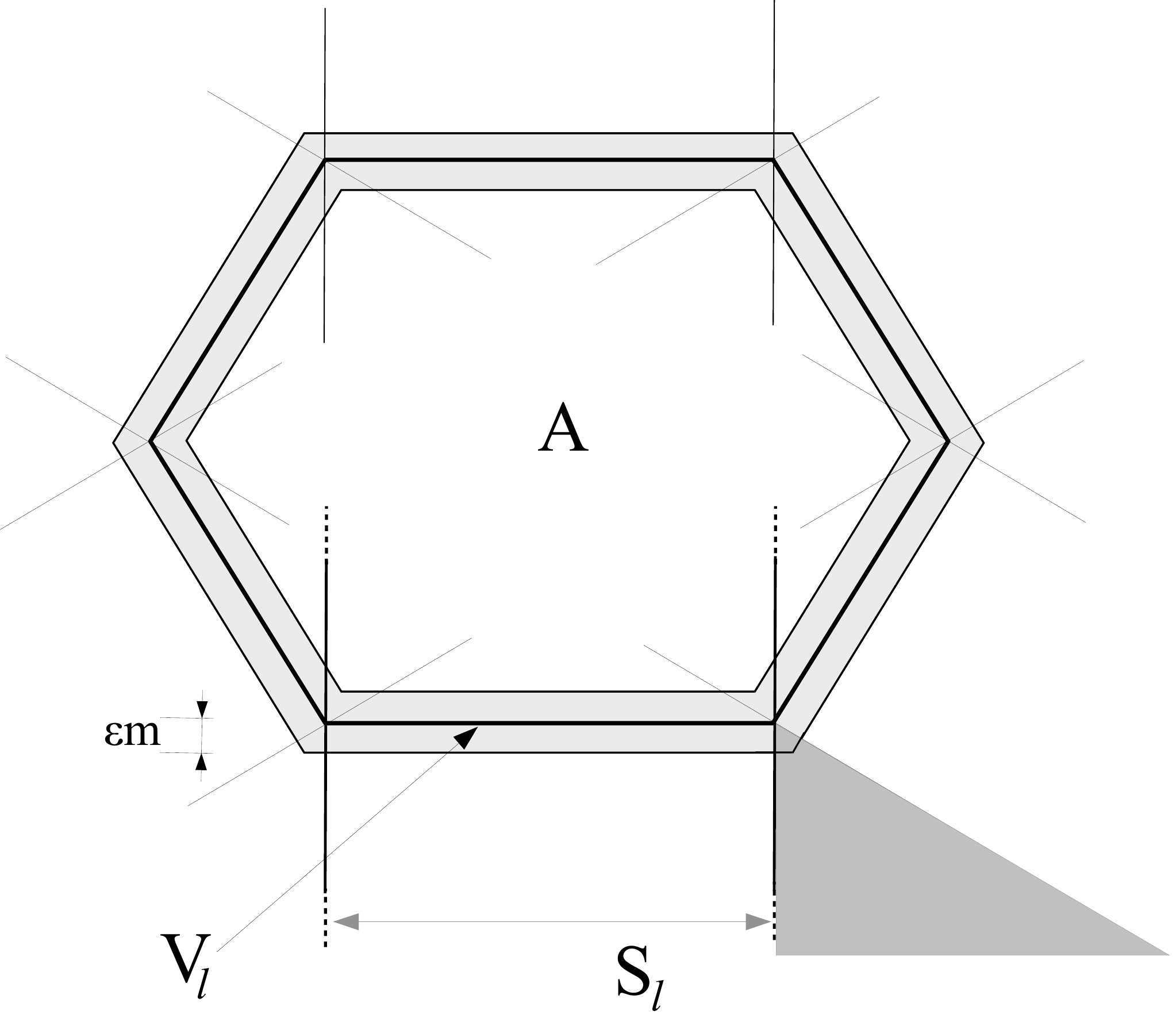}}
\caption{different zones used in estimating the energy.}\label{estima}
   \end{figure}
   The estimates will be separated into different zones: stripes perpendicular to each phase, which we further subdivide into contributions close to (light grey zone in Fig.~\ref{estima}) or far from the boundary, and its complement that we further parameterize in terms of the $(n-2)$-dimensional polytopes of $\partial A$ (in the two-dimensional picture in Fig.~\ref{estima} this reduces to the contribution in each angle as the one highlighted in dark grey).   

For fixed $l\in\I$ we have that
\begin{eqnarray}\label{upperbound-primo}\nonumber
\sum_{i\in\J_{l,\e}}\e^{n-1}\phi(\{v_{\e}^{i+j}\}_{j})&\leq &\sum_{i\in\J_{l,\e}}\e^{n-1}c_{\alpha(i)}\\
\nonumber
&\leq& \sum_{\alpha=1}^{+\infty}\e^{n-1}\# \{i\in\J_{l,\e}:\,\alpha_{\e}(i)=\alpha\}c_{\alpha}\\&\leq& 
C(\H^{n-1}(V_{l})+o_{\e}(1))\sum_{\alpha=1}^{+\infty}c_{\alpha}
\end{eqnarray}
We estimate the second term in \eqref{upperbound} as
\begin{eqnarray}\label{upperbound 2}		\nonumber
\sum_{i\not\in\bigcup_{l\in\I}\J_{l,\e}}\e^{n-1}\phi(\{v_{\e}^{i+j}\}_{j})&\leq&\sum_{l\neq l'} \sum_{i\in {\mathcal C}^{m}_{l,l',\e}} \e^{n-1}\phi(\{v_{\e}^{i+j}\}_{j})\\
&&+\sum_{l\neq l'} \sum_{i\in \Z^{n}\setminus{\mathcal C}^{m}_{l,l',\e}} \e^{n-1}c_{\alpha_{\e}(i)},
\end{eqnarray}
where 
$$
{\mathcal C}^{m}_{l,l',\e}=\{i\in \Z^{n}:\, {\rm dist}_{\infty}(\e i, V_{l}\cap V_{l'})\leq \e m\}.
$$

For the first sum we use that $\phi\leq L$ to get that
\begin{eqnarray}\nonumber
\sum_{l\neq l'} \sum_{i\in {\mathcal C}^{m}_{l,l',\e}} \e^{n-1}\phi(\{v_{\e}^{i+j}\}_{j})&\leq& \sum_{l\neq l'} CL\e^{n-1}\#{\mathcal C}^{m}_{l,l',\e}\\\nonumber
&\leq&\sum_{l\neq l'} \e m^{2}CL(\H^{n-2}(V_{l}\cap V_{l'})+o_{\e m}(1) )\\&\leq& C\e m^{2}.
\end{eqnarray}
For the second sum we need to define for all $i\in\Z^{n}$ 
\begin{equation}
R_{\e}(i)=\sup\{R:\, (\e i+\e RQ)\cap \partial A=\emptyset\}.
\end{equation}
We note that, since $\bigcup_{l\in\I}\J_{l,\e}=\bigcup_{l\in\I}\frac{1}{\e}S_{l}$, for all $i\not\in\bigcup_{l\in\I}\J_{l,\e}$ we have that 
$$
R_{\e}(i)={\rm dist}_{\infty}(\e i, \partial V_{l}\cap\partial V_{l'})
$$
for some $l,l'\in\I$, $l\neq l'$. We set

\begin{equation}
\I_{l,l',\e}=\{i\not\in\bigcup_{l\in\I}\J_{l,\e}:\, R_{\e}(i)={\rm dist}_{\infty}(\e i, V_{l}\cap V_{l'})\}
\end{equation}
Note that for such $i$ we have that $\lfloor R_{\e}(i)\rfloor\leq \alpha_{\e}(i)\leq \lfloor R_{\e}(i)\rfloor+1$ so that
\begin{eqnarray}\nonumber
\#\{i\in \I_{l,l',\e}:\, \alpha_{\e}(i)=\alpha\}&\leq& C \frac{1}{\e^{n-1}}\H^{n-1}\{x:\, {\rm dist}(x, V_{l}\cap V_{l'})=\e(\alpha+1)\}\\\nonumber
&\leq&\frac{1}{\e^{n-1}} C({\rm diam} (V_{l}\cap V_{l'})+\e(\alpha+1))^{n-1}\\&\leq& \frac{1}{\e^{n-1}} C({\rm diam} (A)+\e(\alpha+1))^{n-1}
\end{eqnarray}
and
\begin{eqnarray}\nonumber
\sum_{i\in \Z^{n}\setminus{\mathcal C}^{m}_{l,l',\e}} \e^{n-1}c_{\alpha_{\e}(i)}&\leq& C\sum_{\alpha\geq m} ({\rm diam} (A)+\e(\alpha+1))^{n-1}c_{\alpha}\\
&\leq& C\Bigl({{\rm diam}\,A\over m}\Bigr)^{n-1}\sum_{\alpha\geq m} \alpha^{n-1}c_{\alpha}
\end{eqnarray}
From the estimates above we deduce that
\begin{eqnarray}
F_{\e}(v_{\e})&\leq&C\sum_{l\in\I}\H^{n-1}(V_{l})+C\e m^{2}+C\sum_{\alpha\geq m} \alpha^{n-1}c_{\alpha}.
\end{eqnarray}
Taking the $\limsup$ as $\e\to 0$ and by the arbitrariness of $m$ we obtain the desired estimate \eqref{upperbound-statement}.

\bigskip

For a general Lipschitz open set $\Omega$ we may suppose that $A$ is such that
$$
\H^{n-1}(\partial A\cap \Omega^{\eta})= \H^{n-1}(\partial A\cap \Omega)+ o_{\eta}(1),
$$
where $\Omega^{\eta}=\{x\in\R^{n}: {\rm dist}\,(x,\Omega)<\eta\}$. We may then define $v_{\e}$ as above,
and estimate, as in \eqref{upperbound},
\begin{eqnarray}\label{upperbound-omega}
F_{\e}(v_{\e},\Omega)=\sum_{l\in\I}\sum_{i\in\J_{l,\e}\cap{1\over\e}\Omega}\e^{n-1}\phi(\{v_{\e}^{i+j}\}_{j})+\sum_{i\in{1\over\e}\Omega\setminus\bigcup_{l\in\I}\J_{l,\e}}\e^{n-1}\phi(\{v_{\e}^{i+j}\}_{j}).
\end{eqnarray}
The second sum can be estimated by the second term in \eqref{upperbound}, giving a negligible contribution as $\e\to0$.
As for the first sum, we note that in $\J_{l,\e}\cap{1\over\e}\Omega$ only indices $i\in \Z^{n}\cap {1\over\e}S_{l}$ with $\e i$ with distance larger than $\eta$ from  $V_{l}\setminus \Omega_{\eta}$ are present, so that $\alpha_{\e}(i)\ge C\eta/\e$ for some positive constant. As a result, proceeding as in \eqref{upperbound-primo},  for fixed $m\in\N$ we obtain
\begin{eqnarray}\nonumber
&&
\hskip-2cm \sum_{l\in\I}\sum_{i\in\J_{l,\e}\cap{1\over\e}\Omega}\e^{n-1}\phi(\{v_{\e}^{i+j}\}_{j})\\&\le &
C\Biggl(\H^{n-1}(\Omega^\eta\cap \partial A)\sum_{\alpha\ge 1}c_{\alpha}+\H^{n-1}(\partial A\setminus \Omega^\eta)\sum_{\alpha\ge m}c_{\alpha}\Biggr)
\end{eqnarray}

In the case of a partition into $K$ polyhedral sets the proof is the same upon defining $v_{\e}^{i}=v_{l}$ for $\e i\in A_{l}$, with only a more complicated notation.

Finally, we may remove the assumption that the partition be composed by polyhedral sets using the approximation argument of Remark \ref{appro-polse}.
\end{proof}

\begin{theorem}[compactness and integral representation]\label{teo:comp-intrep}
From every sequence $\{\e_j\}$ we can extract a subsequence (still denoted by $\{\e_j\}$) such that there exists the $\Gamma$-limit of $F_{\e_j}(\cdot,\Omega)$ for every $\Omega$ Lipschitz open set, denoted by $F_0(\cdot,\Omega)$,
and there exists a function $\varphi:\{1,\dots,K\}^{2}\times S^{n-1}\to [0,+\infty)$ such that
\begin{equation}
F_0({\mathbb A},\Omega)=\sum_{l,l'\in \{1,\dots,K\}}\int_{\Omega\cap\partial A_l\cap\partial A_{l'}}\varphi(l,l',\nu_{l,l'})d\H^{n-1}
\end{equation}
for all partitions ${\mathbb A}$, where $\nu_{l,l'}$ denotes the internal normal to $A_{l'}$ at a point in $\partial A_l\cap\partial A_{l'}$.
\end{theorem}

\begin{proof}
The proof follows the localization method of $\Gamma$-convergence. First, upon extracting a subsequence, we may suppose that $F_{\e_j}(\cdot,V)$ $\Gamma$-converges to $F_0(\cdot,V)$
for all $V$ finite union of $n$-dimensional rectangles with rational vertices, since this is a countable class of sets. 

Given a partition ${\mathbb A}$ for all sets $V$ we define 
\begin{eqnarray*}
F'({\mathbb A},V)=\Gamma\hbox{-}\liminf_j F_{\e_j}({\mathbb A},V),
\qquad
F''({\mathbb A},V)=\Gamma\hbox{-}\limsup_j F_{\e_j}({\mathbb A},V).
\end{eqnarray*}
We note that for $\Omega$ a Lipschitz set we have 
\begin{equation}\label{stiff}
F''({\mathbb A},\Omega)=\sup\{F''({\mathbb A},\Omega'): \Omega'\subset\subset \Omega\}.
\end{equation}
This follows from Lemma \ref{stimaf} and Proposition \ref{upperbound}. Indeed, choosing 
$V=\Omega$, $U=\Omega'$ and $U'\subset\subset\Omega'$, $\{u_{U,{\e_j}}\}$ and $\{u_{V,{\e_j}}\}$ giving $F''({\mathbb A},\Omega')$ and $F''({\mathbb A},\Omega\setminus U')$,
using the $v_{\e_j}$ constructed in the lemma to test $F''({\mathbb A},\Omega)$, we get
$$
F''({\mathbb A},\Omega)\le F''({\mathbb A},\Omega')+F''({\mathbb A},\Omega\setminus U')
$$
by \eqref{stimaf-formula}. By Proposition  \ref{upperbound} the last term can be made arbitrarily small as $\Omega'$ tends to $\Omega$, proving \eqref{stiff}. 

From \eqref{stiff} it follows that the $\Gamma$-limit $F_0({\mathbb A},\Omega)$ exists for all $\Omega$ Lipschitz sets (see \cite{HMI} Theorem 10.3). 

In order to prove the integral representation, we will make use of Theorem 3.1 in \cite{AmBrI}.
To this end it suffices to prove that the set function defined on open sets by
$$
\gamma_{\mathbb A}(V)=\sup\{F''({\mathbb A},V'): V'\subset\subset V\}
$$
satisfies:

(i) $0\le \gamma_{\mathbb A}(V)\le C\sum_{l=1}^K\H^{n-1}(V\cap \partial A_l)$;

(ii) $\gamma_{\mathbb A}$ is the restriction of a Borel measure to the family of open sets;

(iii) $\gamma_{\mathbb A}(V)=\gamma_{\mathbb A'}(V)$ if $|(A_l\triangle A'_{l})\cap V|=0$ for all $l\in\{1,\ldots, K\}$;

(iv) $\mathbb A\mapsto \gamma_{\mathbb A}(V)$ is lower semicontinuous with respect to the $L^1_{\rm loc}$ convergence of sets in the partition;

(v) $\gamma_{\mathbb A}(V+z)= \gamma_{\mathbb A_z}(V)$, where $(A_z)_l= A_l-z$.

Note that in the definition of $\gamma$ we can restrict the supremum to Lipschitz $U'$. Property (i) directly follows from Proposition \ref{upperbound}. Property (iii) can be obtained from Lemma \ref{stimaf} as follows: we choose $U$, $U'$ and $V'$ Lipschitz sets with 
$$
U'\subset\subset U\subset\subset V'\subset\subset V,
$$ 
and $u_{U,\e}\to \mathbb A$ and $u_{V,\e}\to \mathbb A'$ such that 
\begin{eqnarray}
\limsup_{\e}F_{\e}(u_{U,\e},U)&=&F''(\mathbb A,U),\\
\limsup_{\e}F_{\e}(u_{V,\e},V'\setminus \overline U')&=&F''(\mathbb A',V'\setminus \overline U').
\end{eqnarray}
Observing that $v_{\e}\to \mathbb A'$ and that the last term of the estimate of \eqref{stimaf-formula} are negligible, we obtain
$$
F''(\mathbb A', V')\leq F''(\mathbb A, U)+c\sum_{l=1}^{K}\H^{n-1}(\partial A'_{l}\cap V'\setminus \overline U')
$$
where we have used Proposition \ref{upperbound}. Letting $U'$ invade $V'$ we get
$$
F''(\mathbb A', V')\leq F''(\mathbb A, V')
$$
and hence $\gamma_{\mathbb  A'}(V)\leq \gamma_{\mathbb  A}(V)$. The property (iii) follows by exchanging the role of $\mathbb A$ and $\mathbb A'$.

Property (iv) follows since $\mathbb A\mapsto\gamma_{\mathbb A}$ is the supremum of a family of lower semicontinuous functions by definition.

Property (v) can be proved as follows: if $V'\subset\subset V$ is a Lipschitz set and $u_{\e}\to \mathbb A$ is a recovery sequence in $V'$, we choose $i_{\e}=\e\lfloor\frac{z}{\e}\rfloor$ and define $v_{\e}^{i}=u_{\e}^{i-i_{\e}}$. Observing that $v_{\e}\to \mathbb A_{z}$ we have, for all $V''\subset\subset V'+z$
\begin{equation}
F_{\e}(v_{\e},V'')\leq F_{\e}(v_{\e}, V'+\e i_{\e})=F_{\e}(u_{\e}, V'),
\end{equation}
hence $F''(\mathbb A_{z}, V'')\leq F''(\mathbb A, V')$ from which we deduce that $\gamma_{\mathbb A_{z}}(V)\leq \gamma_{\mathbb A}(V)$. Property $(v)$ follows by a symmetry argument.

To prove (ii) we use the De Giorgi-Letta criterion (see \cite{DGL}) which requires to prove that $\gamma_{\mathbb A}$ satisfies

(a) $\gamma_{\mathbb A}$ is an increasing set function;

(b) $\gamma_{\mathbb A}$ is superadditive on disjoint sets;

{(}c) $\gamma_{\mathbb A}$ is inner regular;

(d) $\gamma_{\mathbb A}$ is subadditive.

Properties (a) and (c) follow immediately from the definition of $\gamma_{\mathbb A}$. 


To prove (b) note that, since for Lipschitz sets 
$F''=F_0$,  it suffices to show that
$$
F_0(\mathbb A,V)+ F_0(\mathbb A,V')\le F_0(\mathbb A,V\cup V')
$$
when $V$ and $V'$ are Lipschitz open sets with positive distance. 
This inequality follows
immediately using a recovery sequence for the $\Gamma$-limit
on the right-hand side as test functions for the two $\Gamma$-limits
on the left-hand side.


To prove (d) we need to show that, given $Z$ and $W$ two open subsets of $\R^{n}$, it holds
\begin{equation}\label{gamma-sub}
\gamma_{\mathbb A}(Z\cup W)\leq \gamma_{\mathbb A}(Z)+\gamma_{\mathbb A}(W).
\end{equation}
For $Z'\subset\subset Z$ and $W'\subset\subset W$ we make use of Lemma \ref{stimaf} in which we choose $V=Z'\cup W'$, $V'\subset \subset V$, $V'\setminus W'\subset U'\subset\subset U\subset\subset Z'$, $u_{U,\e}$ and $u_{V,\e}$ recovery sequences for the $\Gamma\hbox{-}\limsup$ for $\mathbb A$ in $Z'$ and $W'$, respectively. By \eqref{stimaf-formula} we have
\begin{equation}
F_{\e}(v_{\e}, V')\leq F_{\e}(u_{U,\e}, Z')+F_{\e}(u_{V,\e}, W')+o_{\e}(1)+o_{R}(1).
\end{equation} 
Observing that $v_{\e}\to \mathbb A$ we obtain that
\begin{equation}
F''(\mathbb A, V')\leq F''(\mathbb A, Z')+F''(\mathbb A, W'),
\end{equation} 
hence \eqref{gamma-sub} by definition of $\gamma_{\mathbb A}$.
\end{proof}

\subsection{An extension: a general compactness result}
It is possible to extend the statement of Theorem \ref{teo:comp-intrep} to energy densities depending on $\e$ if we assume that (H1) and (H2) be satisfied uniformly with respect to $\e$ and if (H3) is modified as follows:\\

(H3) (mild non-locality) given $z,w\in{\mathcal A}$ and $m\in\N$ such that $z^j=w^j$ for all $j\in m Q\cap\Z^n$, then 
$$
|\phi_\e(\{w^j\}_{j\in\Z^n})-\phi_\e(\{z^j\}_{j\in\Z^n})|\le c^{\e}_m
$$
where the constants $c^\e_m$ are such that 
$$
\limsup_\e\sum_{m\in\N}c^\e_mm^{n-1}<+\infty
$$
and that for all $\delta>0$ there exists $M_\delta>0$ such that 
\begin{equation}\label{H4}
\limsup_\e\sum_{m>M_\delta}c^\e_mm^{n-1}<\delta.
\end{equation}
The proof the compactness result under this assumption follows exactly the one of Theorem \ref{teo:comp-intrep} applying (\ref{H4}) in the estimate of the remainder terms in Proposition \ref{upperbound}.

Furthermore, the same proof also applies to inhomogeneous energy densities $\phi^i_\e$, provided
that all assumptions are satisfied uniformly in $i$. Note that in this case we need to refer to Theorem 3.2 in \cite{BCP} for the integral representation result, presented there in the more general context of infinite partitions,
highlighting that the limit energy density $\varphi$ also depends on $x$. Since the proof involves only 
notational changes, we do not include the details.

\subsection{Homogenization formula}
In order to characterize the energy density $\varphi$ and to prove that the whole family 
$F_\e$ converges as $\e\to 0$ we first prove a homogenization formula in terms
of minimum problems related to $F$.

\begin{proposition}\label{proposition:existence}
For $\d, T>0$, $l,l'\in\{1,2,\dots,K\}$ and $\nu\in{\mathcal S}^{n-1}$ we set
\begin{equation*}
m_\d^{l,l'}(TQ_\nu):=\min\{F(u,TQ_{\nu}),\ u\in{\mathcal B}^{l,l'\!\!,\nu}_{\d}(TQ_\nu)\}.
\end{equation*}
Then the following equality holds
\begin{eqnarray}\label{prop:exists}
\inf_{\delta>0}\liminf_{T\to +\infty}\frac{1}{T^{n-1}}m_\d^{l,l'}(TQ_\nu)
=\inf_{\delta>0}\limsup_{T\to +\infty}\frac{1}{T^{n-1}}m_\d^{l,l'}(TQ_\nu).
\end{eqnarray}
\end{proposition}
\begin{proof}
Given $\nu\in S^{n-1}$ we let $\{\nu^{\perp}_{1},\nu^{\perp}_{2},\dots,\nu^{\perp}_{n-1}\}$ be
an orthonormal basis of $\Pi_{\nu}$. For all $i=(i_{1},i_{2},\dots,i_{n-1})\in (T+\sqrt{n})\Z^{n-1}$ we set $z_{i}=\sum_{k=1}^{n-1}i_{k}\nu^{\perp}_{k}$.
For $\delta>0$ and $\beta\in\N$ we take $T,S>0$ such that $S>>T$ and $\delta T>4\beta$ and we set
\begin{equation}
Z_{\d}(T,S):=\{i\in(T+\sqrt{n})\Z^{n-1}:\, \lfloor z_{i}\rfloor+TQ_{\nu}\subseteq(1-\d)SQ_{\nu}\}
\end{equation}
Given $l,l'\in\{1,2,\dots,K\}$ we take
$u_{T}\in{\mathcal B}^{l,l'\!\!,\nu}_{\d}(TQ_\nu)$ such that $F(u_{T},TQ_{\nu})=m_\d^{l,l'}(TQ_\nu)$.
We now define $u_{S}\in{\mathcal B}^{l,l'\!\!,\nu}_{\d}(SQ_\nu)$ as
\begin{eqnarray}
u_{S}^{j}=
\begin{cases}
u_{T}^{j-\lfloor z_{i}\rfloor} & \hbox{if } j\in \lfloor z_{i}\rfloor+TQ_{\nu} \hbox{ and } i\in Z_{\d}(T,S),\\
u^{j}_{l,l'\!\!,\nu}& \hbox{otherwise.}
\end{cases}
\end{eqnarray}
In order to estimate $F(u_{S},SQ_{\nu})$ we rewrite it as the sum of three terms
\begin{eqnarray*}
F(u_{S},SQ_{\nu})&=&\sum_{i\in Z_{\d}(T,S)}F(u_{S},\lfloor z_{i}\rfloor+(T-\sqrt{n}\beta)Q_{\nu})\\
&&+\sum_{i\in  SQ_\nu\cap P_{\d,\beta}(T,S)}\phi(\{u_{S}^{i+j}\}_{j})+\sum_{i\in SQ_\nu\setminus P_{\d,\beta}(T,S)}\phi(\{u_{S}^{i+j}\}_{j})\\
&=:&\Sigma_{1}+\Sigma_{2}+\Sigma_{3},
\end{eqnarray*}
where we have set 
$$
P_{\d,\beta}(T,S):=\Big\{i\in \Z^{n}\setminus \bigcup_{j\in Z_{\d}(T,S)}
(\lfloor z_{j}\rfloor+(T-\sqrt{n}\beta)Q_{\nu})):\, {\rm dist}(i,\Pi_{\nu})\leq \beta\Big\}.
$$
For all $i\in Z_{\d}(T,S)$ we have that 
\begin{eqnarray}\nonumber
F(u_{S},\lfloor z_{i}\rfloor+(T-\sqrt{n}\beta)Q_{\nu})&\leq& F(u_{T},(T-\sqrt{n}\beta)Q_{\nu})+C\sum_{\alpha\geq \beta}c_{\alpha}T^{n-1}\\
&\leq& F(u_{T},TQ_{\nu})+C\sum_{\alpha\geq \beta}c_{\alpha}T^{n-1},
\end{eqnarray}
where in the first inequality we have used the energy estimate in (H3) observing that for all $i\in Z_{\d}(T,S)$ the functions $u_{S}$ and $u_{T}$ agree on the cubes $\lfloor z_{i}\rfloor+TQ_{\nu}$, while in the second inequality we have used the non-negativity of $\phi$.
By the previous estimate, observing that 
$$
\#Z_{\d}(T,S)\leq (1-\d)\frac{S^{n-1}}{T^{n-1}},
$$
we get 
\begin{equation}\label{Sigma-1}
\Sigma_{1}\leq (1-\d)\frac{S^{n-1}}{T^{n-1}}F(u_{T},TQ_{\nu})+CS^{n-1}\sum_{\alpha\geq \beta}c_{\alpha}\alpha^{n-1}.
\end{equation}
Since $\phi\leq L$ we estimate $\Sigma_{2}$ as 
\begin{equation}\label{Sigma-2}
\Sigma_{2}\leq CL\beta\d S^{n-1}+CL(\beta T^{n-2})\frac{S^{n-1}}{T^{n-1}}.
\end{equation}
In order to estimate $\Sigma_{3}$ we observe that for all $i\in P_{\d,\beta}(T,S)$ there exists $\alpha(i)\geq \beta$ such that the function $u_{S}$ agrees with $v_{l}$ or with $v_{l'}$ on all cubes $i+\alpha(i)Q$ and that $\#\{i\in P_{\d,\beta}(T,S):\, \alpha(i)=\alpha\}\leq C S^{n-1}\alpha^{n-1}$. 
Therefore we have that
\begin{equation}\label{Sigma-3}
\Sigma_{3}\leq CS^{n-1}\sum_{\alpha\geq \beta}c_{\alpha}\alpha^{n-1}.
\end{equation}
Gathering \eqref{Sigma-1}, \eqref{Sigma-2} and \eqref{Sigma-3} we have
\begin{equation*}
\frac{1}{S^{n-1}}m_\d^{l,l'}(SQ_\nu)\leq \frac{1}{T^{n-1}}m_\d^{l,l'}(TQ_\nu)+C\sum_{\alpha\geq \beta}c_{\alpha}\alpha^{n-1}+CL\beta\Big(\d+\frac{1}{T}\Big),
\end{equation*}
which further implies that
\begin{equation*}
\limsup_{S\to +\infty}\frac{1}{S^{n-1}}m_\d^{l,l'}(SQ_\nu)\leq \liminf_{T\to +\infty}\frac{1}{T^{n-1}}m_\d^{l,l'}(TQ_\nu)+C\Big(\sum_{\alpha\geq \beta}c_{\alpha}\alpha^{n-1}+\beta\d\Big).
\end{equation*}
Observing that $\d\mapsto\liminf_{T}\frac{1}{T^{n-1}}m_\d^{l,l'}(TQ_\nu)$ is an increasing function, passing to the infimum on $\d$ in the previous inequality we get
\begin{equation*}
\inf_{\d>0}\limsup_{S\to +\infty}\frac{1}{S^{n-1}}m_\d^{l,l'}(SQ_\nu)\leq \inf_{\d>0}\liminf_{T\to +\infty}\frac{1}{T^{n-1}}m_\d^{l,l'}(TQ_\nu)+C\Big(\sum_{\alpha\geq \beta}c_{\alpha}\alpha^{n-1}\Big).
\end{equation*}
Equality \eqref{prop:exists} now follows by passing to the limit as $\beta$ diverges in the previous inequality since by (H3) $\lim_{\beta}\sum_{\alpha\geq \beta}c_{\alpha}\alpha^{n-1}=0$.
\end{proof}

\subsection{Boundary-value problems}
In this section we consider minimum problems with Dirichlet-type boundary conditions for $F_{\e}$. 
It will be convenient to impose the boundary condition in a neighborhood of the boundary. To this end in what follows, for $\Omega\subset\R^{n}$ a bounded Lipschitz set and $\delta>0$ we denote by $\Omega_{\delta}$ the $\delta$-thick boundary of $\Omega$ defined as in \eqref{boundary} with $\delta$ in place of $\eta$.

\begin{proposition}\label{prop:bound}
Let $\Omega\subset\R^{n}$ be a bounded Lipschitz set and let ${\mathbb A}_{0}$ be a partition of $\R^{n}$ in sets of finite perimeter. 
Suppose that $F_{\e}$ satisfy {\rm(H1)}--{\rm(H3)} and that (upon extracting a subsequence) there exists 
$$
F_0({\mathbb A},\Omega)=\Gamma\hbox{-}\lim_{\e\to 0}F_{\e}(u,\Omega).
$$ 
Let $u_{0,\e}\in\mathcal A$ be such that $u_{0,\e}\to{\mathbb A}_{0}$ and that
\begin{eqnarray}\label{u0buone}
\inf_{\delta}\limsup_{\e\to 0}F_{\e}(u_{0,\e},\Omega\setminus\overline \Omega_{2\delta})=0.
\end{eqnarray}
Then, setting 
\begin{eqnarray*}
m_{\e,\delta}&=&\min\Bigl\{F_{\e}(u,\Omega):\, u^{i}=u^{i}_{0,\e},\,\, i\in\Z^{n}\setminus{1\over\e}\Omega_{\delta} \Bigr\},\\
m_{0,\delta}&=&\min\{F_0({\mathbb A},\Omega):\,{\mathbb A}={\mathbb A}_{0} \hbox{ on }\R^{n}\setminus\Omega_{\delta}\},
\end{eqnarray*}
it holds
\begin{eqnarray}\label{bound-est}
\inf_{\delta}\limsup_{\e\to 0}m_{\e,\delta}\leq\inf_{\delta} m_{0,\delta}\leq\inf_{\delta}\liminf_{\e\to 0}m_{\e,\delta}.
\end{eqnarray}
\end{proposition}
\begin{proof}
Let $u_{\delta,\e}$ be a minimizer for $m_{\e,\delta}$. By Theorem \ref{teo:comp} we have that $u_{\delta,\e}\to{\mathbb A}_{\delta}$ with ${\mathbb A}_{\delta}=\mathbb A$ on $\R^{n}\setminus \Omega_{\delta}$. Using the lower semicontinuity inequality  we have 
\begin{equation*}
m_{0,\delta}\leq F_0({\mathbb A}_{\delta},\Omega)\leq\liminf_{\e\to 0}F_{\e}(u_{\delta,\e},\Omega)=\liminf_{\e\to 0}m_{\e,\delta},
\end{equation*}
by which, taking the infimum over $\delta$ one obtains the second inequality in \eqref{bound-est}.\\

Fix $\mathbb A$ a minimizer in $m_{0,\delta}$ and let $u_{\e}$ be a recovery sequence for $F_0$ at $\mathbb A$.
By Lemma \ref{stimaf} with $U'=\Omega_{2\delta}$, $U=\Omega_{\delta}$, $V=\Omega$, $u_{U,\e}=u_{\e}$ and $u_{V,\e}=u_{0,\e}$, there exists $v_{\e}$ such that $v_{\e}=u_{\e}$ on $\Z^{n}\cap\frac{1}{\e}\Omega_{2\delta}$ , $v_{\e}=u_{0,\e}$ on $\Z^{n}\setminus \frac{1}{\e}\Omega_{\delta}$, for $V'=V=\Omega$,
\begin{eqnarray}
\nonumber F_{\e}(v_{\e},\Omega)&\leq& F_{\e}(u_{\e},\Omega)+F_{\e}(u_{0,\e},\Omega\setminus\overline\Omega_{2\delta})\\
&&+O(\e)+o_{R}(1)+CR^{n+1}\|p(u_{\e})-p(u_{0,\e})\|_{L^{1}(\Omega_{\delta}\setminus\Omega_{2\delta})}
\end{eqnarray}
As a consequence
\begin{eqnarray*}
\nonumber\limsup_{\e}F_{\e}(v_{\e},\Omega)&\leq& m_{0,\delta}+\limsup_{\e}F_{\e}(u_{0,\e},\Omega\setminus\overline\Omega_{2\delta})\\
&&+o_{R}(1)+CR^{n+1}\limsup_{\e}\|p(u_{\e})-p(u_{0,\e})\|_{L^{1}(\Omega_{\delta}\setminus\Omega_{2\delta})}.
\end{eqnarray*}
Noting that the applications $\delta\mapsto \limsup_{\e}\|p(u_{\e})-p(u_{0,\e})\|_{L^{1}(\Omega_{\delta}\setminus\Omega_{2\delta})}$ and 
$\delta\mapsto\limsup_{\e}F_{\e}(u_{0,\e},\Omega\setminus\overline\Omega_{2\delta})$ are increasing, taking the infimum over $\delta$ and using \eqref{u0buone} we finally obtain the first inequality in \eqref{bound-est}.
\end{proof}

\begin{remark}\rm
Due to the oscillating nature of the ground states it is not possible to give a simple
statement for general boundary value problems, both of Dirichlet and Neumann type.
This is evident even in a one-dimensional setting for `parity' reasons.

As a simple example, we can take the antiferromagnetic nearest-neighbor spin energy
$\phi(\{u^j\}_j)= (u^0+u^1)^2$ and Dirichlet boundary conditions; e.g.,  $u^0=u^N=1$.
Then the minimizer depends on the parity of $N$, which may force the appearance of an interface.

A similar parity issue applies for free boundary conditions (i.e., no conditions on the extreme values $u^0$ and $u^N$) for
the next-to-nearest antiferromagnetic interactions of Example \ref{mapbnn}. Then 
no interface appears but the minimizer coincides with one of the ground states, which one depending on $N$.
\end{remark}

\subsection{Conclusion of the proof of Theorem \ref{teo:Gamma}}
In this section we prove Theorem \ref{teo:Gamma} collecting the compactness and integral representation results of Theorem \ref{teo:comp-intrep}, the homogenization formula proven in Proposition \ref{proposition:existence} and the convergence of boundary value problems in Proposition \ref{prop:bound}.
\begin{proof}
Let $F_0$ and $\varphi$ be given by Theorem \ref{teo:comp-intrep}.
In particular we have that $\varphi$ is a $BV$-elliptic integrand \cite{AmBrI}; 
i.e., the two-set partition ${\mathbb A}_{l,l',\nu}$, given by $(A_{1},A_{2},\dots A_{k})$ such that $A_{l}=\{x\in \R^{n}:\, (x,\nu)>0\}$, $A_{l'}=\{x\in \R^{n}:\, (x,\nu)<0\}$ and $A_{j}=\emptyset$ for $j\not\in\{l,l'\}$, is minimal among all its perturbations with compact support. In particular, for $\delta>0$ sufficiently small
\begin{equation}
\varphi(l,l',\nu)=\min\{ F_0(\mathbb A,Q_{\nu}):\,{\mathbb A}={\mathbb A}_{l,l',\nu}\hbox{ on } Q_{\nu}\setminus(1-\delta)\overline Q_{\nu}\},
\end{equation}
since $F({\mathbb A}_{l,l',\nu},Q_{\nu})= \varphi(l,l',\nu)$.
By Proposition \ref{prop:bound}  we then deduce that $\varphi$ is given by the asymptotic formulas in Proposition \ref{proposition:existence}, and in particular it does not depend on the
subsequence $\{\e_j\}$. This proves \eqref{teo:Gammalim-for} and \eqref{teo:Gamma-lim}.
\end{proof}

\section{Examples: textures for spin systems}
In this section we include some examples that illustrate the use of Theorem \ref{teo:Gamma}
to describe the behavior of spin systems; i.e., when $X=\{-1,+1\}$. The energy densities that we are going to consider are sums of pair interactions, which can always be distinguished between ferromagnetic (minimized by spins of the same sign) and antiferromagnetic (minimized by spins of opposite sign) pair interactions. We will usually write such pair interactions as $(u^i-u^j)^2$ and $(u^i+u^j)^2$, respectively.

We will include some pictorial description of interactions and ground states. To that end
the values $1$ and $-1$ will be represented by white and black circles, and 
(where this is relevant) ferromagnetic and antiferromagnetic interactions will be 
represented by a single and a double segment, respectively. Finally, when we 
will wish to highlight frustration of an interaction (i.e., that such interaction is not minimized)
then we will use dashed lines.

We note that in all the examples interactions are of finite range, so that (H3) is trivially satisfied.
The other two assumptions are verified by minimizing the energy densities and noting that minimizers are periodic. 

\begin{example}[anti-phase boundaries for antiferromagnetic nearest-neighbor interactions]\label{apban}\rm
Beside the usual ferromagnetic energies, the simplest example is that of antiferromagnetic nearest-neighbor interactions. In order to use Theorem \ref{teo:Gamma}, antiferromagnetic interactions must be rewritten so as to satisfy conditions (H1)--(H3); e.g., taking
\begin{equation}\label{antien}
\phi(\{u^j\}_j)=\sum_{k=1}^n (u^{e_k}+u^0)^2 =2\sum_{k=1}^n (1+u^{e_k}u^0).
\end{equation}\
\begin{figure}[h!]
\centerline{\includegraphics [width=1.5in]{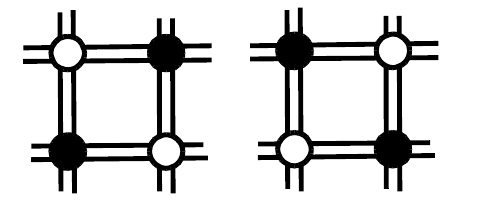}}
\caption{antiferromagnetic alternating ground states}\label{alternating}
   \end{figure}
The two ground states are then the alternating states $v_1$ and $v_2$ 
of period $2$ given by $v_1^i=(-1)^{|i|_1}$ and $v_2=-v_1$, represented in the periodicity cell in dimension $2$ in Fig.~\ref{alternating}.
Note that $v_1$ differs from $v_2$ by a unit translation, so that locally they represent the same 
microscopic pattern. In this case (when the two states on both sides of an interface differ by a translation) we talk about an {\em anti-phase boundary}. 

Note that in this case we have a complete equivalence with a nearest-neighbor ferromagnetic interaction energy by the change of variables $w^i=(-1)^{|i|_1}u^i$, so that the computation of the  
homogenized energy density gives $\varphi(\nu)= 8\|\nu\|_1$ (the constant $8$ due to the normalization in (\ref{antien})) (see also \cite{ABC}).
\end{example}

\begin{example}[modulated anti-phase boundaries for next-to-nearest one-dimensional antiferromagnetic interactions]\label{mapbnn}
\rm 
In dimension one, we consider interactions of the form
\begin{equation}
\phi(\{u^j\}_j)= (u^1+u^{-1})^2+ c(u^1u^0+u^0u^{-1}).
\end{equation}
If $c=0$ this represents an interaction energy between next-to-nearest neighbors only,
which gives decoupled alternating ground states on $2\ZZ$ and $2\ZZ+1$. 
\begin{figure}[h!]
\centerline{\includegraphics [width=3.5in]{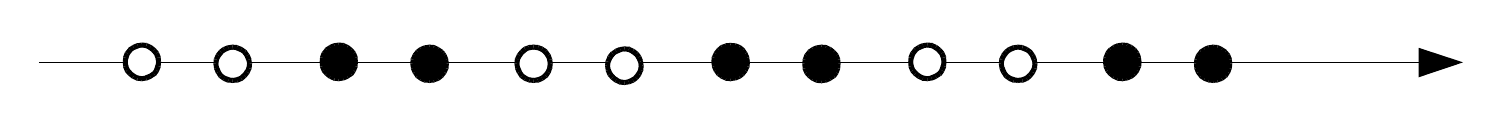}}
\caption{four-periodic ground state}\label{1D4P}
   \end{figure}The ground states are the same if $|c|<2$. In terms of functions defined in $\ZZ$ we have four 
$4$-periodic ground states of the form
$v_k^i = v^{i+k}$ ($k = 1, 2, 3,4 $), where $v$ (equal to $v_4$) is given by
$$
v^i = \begin{cases}
-1 & \text{if } i = 1, 2\\
1 & \text{if } i = 3, 4
\end{cases}
$$
and is represented in Fig.~\ref{1D4P}.

The homogenized energy density of a transition between $v_l$ and $v_{l'}$, 
is obtained by first solving a simplified minimal-transition problem
\begin{eqnarray*}
\psi(k)&=&\min\Bigl\{\sum_{i=-\infty}^{+\infty}\Bigl( (u^{i-1}+u^{i+1})^2+c(u^iu^{i-1} + u^{i+1}u^i) \Bigr):\\
&&\qquad
u^i=v^i\hbox{ for } i\le -4, 
u^i=v_k^i\hbox{ for } i\ge 4\Bigr\}
\end{eqnarray*}
for $k\in\{1,2,3\}$. 
This value is computed over a finite set of states, and is
$$
\psi(k)=\begin{cases} 2\min\{6-3c, 2+c\}& \text{if }k=1\\
8-4|c| & \text{if }k=2\\
2\min\{6+3c, 2-c\}& \text{if } k=3.
\end{cases}
$$
We then have 
$$
\varphi(l,l')=\psi(|l-l'|).
$$
The minima in the definition of $\psi$ highlight the possibility of different types of interfacial structures between two modulated phases. 
\begin{figure}[h!]
\centerline{\includegraphics [width=3.5in]{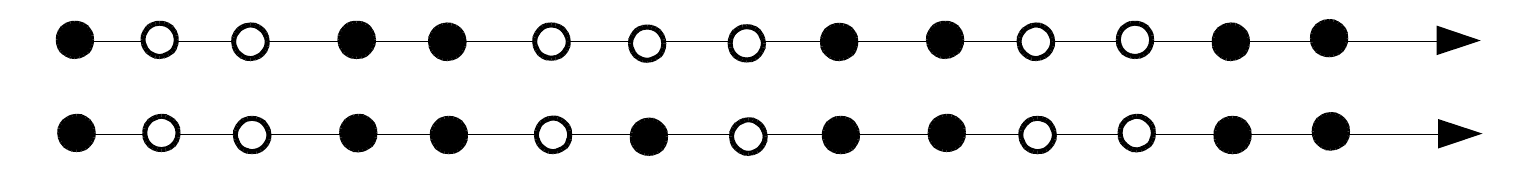}}
\caption{minimal interfacial microstructures}\label{1D4PI}
   \end{figure}
The two possibilities (depending on the value of $c$)
for $k=1$ are represented in Fig.~\ref{1D4PI}.
\end{example}

\begin{example}[phase and anti-phase boundaries for two-dimensional next-to\-nearest antiferromagnetic interactions]\label{ABC_ex}\rm
We briefly rewrite an example treated in detail in \cite{ABC} in our context, involving next-to-nearest antiferromagnetic interactions. In this two-dimensional example we consider
\begin{eqnarray}\label{nn-antien}\nonumber
\phi(\{u^j\}_j)&=&(u^{(1,1)}+u^{(0,0)})^2+ (u^{(1,-1)}+u^{(0,0)})^2\\
&&+(u^{(-1,1)}+u^{(0,0)})^2+ (u^{(-1,-1)}+u^{(0,0)})^2.
\end{eqnarray}
\begin{figure}[h!]
\centerline{\includegraphics [width=1.5in]{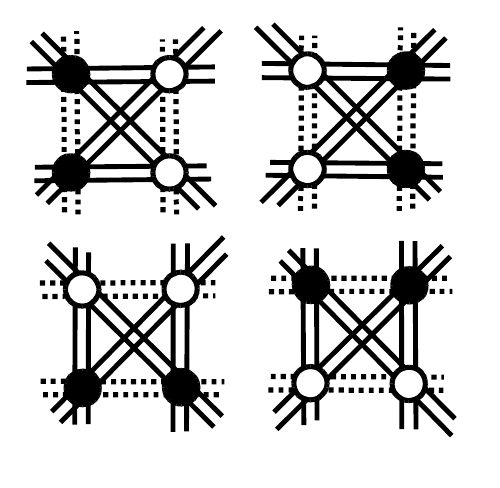}}
\caption{antiferromagnetic `stripe-patterned' ground states}\label{nn-alter}
   \end{figure}
In this way we have nearest-neighbor antiferromagnetic interactions between points of two disjoint square lattices (the one with $i_1+i_2$ odd and the one with $i_1+i_2$ even), on each of which we have two alternating ground states as described in Example \ref{apban} above. Note that,
as in Example \ref{mapbnn}, those two square lattices can be 
coupled by adding a nearest-neighbor term of the form
\begin{equation}\label{nn-cont}
c \Bigl((u^{(1,0)}+u^{(0,0)})^2+(u^{(0,1)}+u^{(0,0)})^2\Bigr)
\end{equation}
 without changing the ground states if $c$ is sufficiently small (both positive or negative). The four two-periodic ground states are represented in Fig.~\ref{nn-alter}, where the dashed lines represent the frustrated nearest-neighbor interactions.

Note that we have both anti-phase boundaries (between ground states differing by a translation) and phase boundaries (between two states given by `stripes' in different directions), which are reflected in the different form of the corresponding homogenized energy densities, for whose computation we refer to \cite{ABC}. If the constant $c$ in the nearest-neighbor contribution \eqref{nn-cont} is large then (after a renormalization of the energy in order to have ground states of zero energy) 
this term dominates, and the ground states are either alternating if $c$ is positive or constant if $c$ is negative.
\end{example}

\begin{example}[multiple phase and anti-phase boundaries for two-dimensional long-range antiferromagnetic interactions]\label{mpapb}\rm
We now extend Example \ref{mapbnn} to dimension two by considering 
 \begin{equation}
\phi(\{u^j\}_j)= (u^{(1,0)}+u^{(-1,0)})^2+  (u^{(0,1)}+u^{(0,-1)})^2,
\end{equation}
where we have an antiferromagnetic interaction between all points at distance $2$ on the lattice $\ZZ^2$.

 \begin{figure}[h!]
\centerline{\includegraphics [width=3.5in]{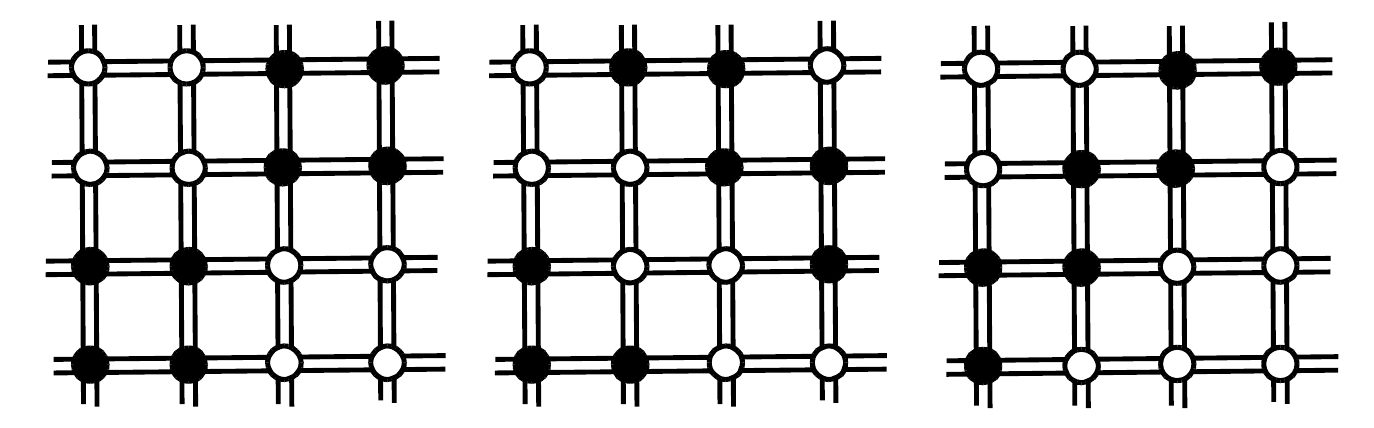}}
\caption{three patterns of ground states}\label{16gs}
   \end{figure}
The resulting energy can be decoupled as the contributions on horizontal and vertical directions, 
noting that on each vertical or horizontal line we have interactions as in Example \ref{mapbnn}.
As a result, minimal states are four-periodic functions with alternating pairs of $+1$ and $-1$ in each direction. Such functions are determined by their values on a square of period two; hence, we have $16$ different ground states, subdivided into three families of modulated phases with different textures: four-periodic checkerboards
(with 8 elements) and stripes at plus or minus 45-degrees, respectively (with 4 elements each). In Fig.~\ref{16gs} we picture the configurations on a period for three different elements, one for each family.

In order to compute the homogenized energy density it is convenient to introduce a vector  parameter, after noticing that the energy can be rewritten as the
sum of four decoupled antiferromagnetic energies on the lattices $2\ZZ^2+ i$ with $i\in\{(0,0),\, (1,0),\, (0,1),\,(1,1)\}$. A ground state $v$ is then described by the vector $y=(v^{(0,0)},v^{(1,0)}, v^{(0,1)},v^{(1,1)})$ in $Y=\{\pm1\}^4$, which plays the role of $l\in\{1,\ldots, K\}$ in Section \ref{Assumption-1}.
Using such a parameter the energy density is simply given by
$$
\varphi(y,y',\nu)= |y-y'|^2\,\|\nu\|_1.
$$
This formula highlights that we have a superposition of the separate energy densities $(y_k-y'_k)^2\|\nu\|_1$ (i.e., $4\|\nu\|_1$ whenever we have an interface for the $k$-th component); the factor $4$ in place of $8$ (see Example \ref{apban}) is due to the fact that each such interaction is considered on the lattice $2\ZZ$. 

As in the previous examples, note that we can add nearest (and also next-to-nearest) neighbor interactions with small coefficients without modifying the patterns of the ground states.

\end{example}

In the next examples we will consider the triangular lattice in $\RR^2$. 
Note that interactions on the triangular lattice can be parameterized on $\ZZ^2$
by suitably identifying first neighbors in the triangular lattice with first and second 
neighbors in the square lattice. Anyhow we will use a parameterization on a unit triangular lattice.

\begin{example}[total frustration for the triangular lattice for antiferromagnetic interaction]\rm
We now check the well-known fact that antiferromagnetic interactions in the triangular lattice
are not described as in Theorem \ref{teo:Gamma}. We may consider the normalized antiferromagnetic energy density $\phi$
as 
 \begin{equation}
\phi(\{u^j\}_j)=u^{(0,0)}u^{(1,0)}+ u^{(1,0)}u^{(1/2,\sqrt3/2)}+u^{(1/2,\sqrt3/2)}u^{(0,0)}+1.
\end{equation}
The corresponding energy is minimized by functions $u$ not taking a constant value on each
unit triangle of the lattice. 

 \begin{figure}[h!]
\centerline{\includegraphics [width=1.5in]{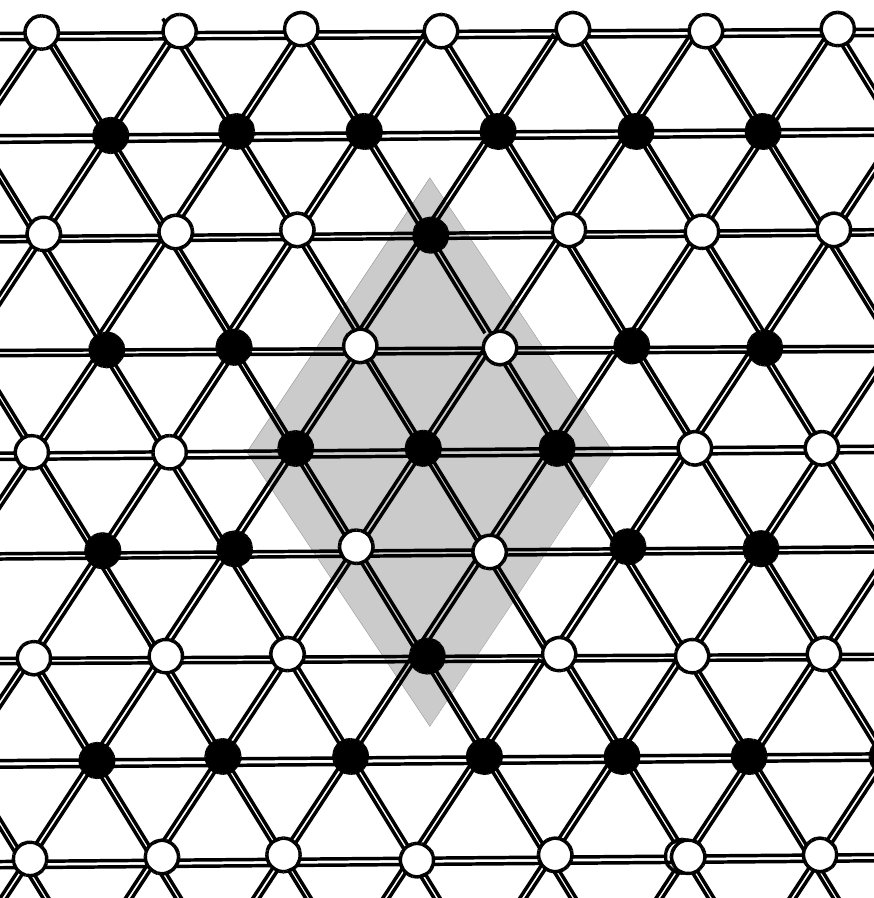}}
\caption{compact-support perturbation of a periodic ground state}\label{rombinc}
   \end{figure}
\end{example}
In order to check that a parameterization with a finite number of ground states is not possible, we show that there are infinitely many distinct periodic ground states. To that end it suffices to exhibit a compact-support perturbation of a periodic ground state. Indeed, we can choose as a periodic ground state the one with alternate horizontal lines of $+1$ and $-1$, and we remark that we can 
exchange the values on those lines on a rhombus as pictured in Fig. \ref{rombinc}.
By placing periodically copies of such a rhombus we can construct minimal states with arbitrary period.

\begin{example}[`stabilization' for the triangular lattice for antiferromagnetic interaction]\rm
We now show that the addition of a (even small) ferromagnetic or antiferromagnetic next-to-nearest neighbor interaction to a system of antiferromagnetic nearest-neighbor interactions 
allows to apply  Theorem \ref{teo:Gamma}. 

We first consider ferromagnetic next-to-nearest neighbor interaction, choosing
 \begin{eqnarray}\nonumber
\phi(\{u^j\}_j)&=&u^{(0,0)}u^{(1,0)}+ u^{(1,0)}u^{(1/2,\sqrt3/2)}+u^{(1/2,\sqrt3/2)}u^{(0,0)}+1\\
&&+c\Bigl((u^{(0,0)}-u^{(0,\sqrt  3)})^2+(u^{(0,0)}-u^{(3/2,-\sqrt  3/2)})^2\nonumber \\
&&+ 
(u^{(0,0)}-u^{(-3/2,-\sqrt  3/2)})^2\Bigr),
\end{eqnarray}
the ferromagnetic part selects ground states with $u$ constant on the three triangular
sub-lattices with triangles of size length $\sqrt 3$.  \begin{figure}[h!]
\centerline{\includegraphics [width=1.2in]{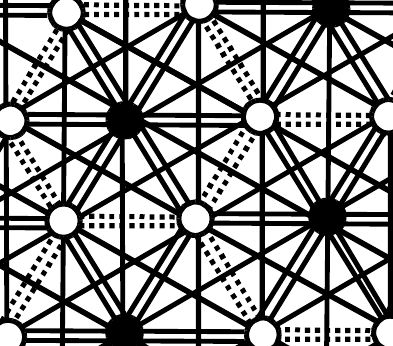}\qquad\qquad\includegraphics [width=1.2in]{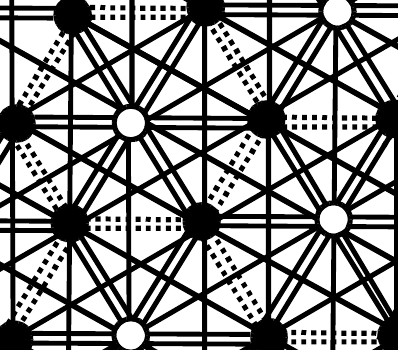}}
\caption{hexagonal patterns in the triangular lattice}\label{hegatri}
   \end{figure}
   As a result, the ground states are the six non-constant functions
which are constant on each of the three lattices. They form two families of modulated phases
with hexagonal symmetries, two elements of which are pictured in Fig.~\ref{hegatri} (the other ones differing by a translation).

 As in Example \ref{mpapb}, in order to compute the homogenized energy density it is convenient to label a ground states $v$ with $(v^{(0,0)}, v^{(1,0)}, v^{(1/2,\sqrt3/2)})\in Y:=\{\pm1\}^3\setminus \{(1,1,1),(-1,-1,-1)\}$. By decoupling the interactions on the three `ferromagnetic' triangular lattices
 one obtains 
 $$
 \varphi(y,y',\nu)= c|y-y'|^2 \,\varphi_{\rm hex}(\nu),
 $$
 where $\varphi_{\rm hex}$ denotes the homogenized energy for the ferromagnetic interactions on one of the three lattices (see \cite{ABC}).

  \begin{figure}[h!]
\centerline{\includegraphics [width=1.2in]{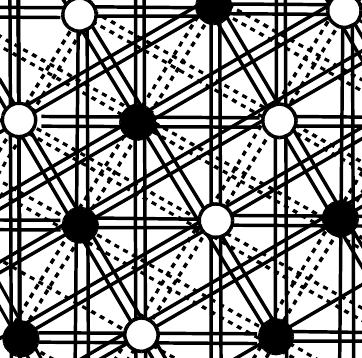}}
\caption{striped patterns in the triangular lattice}\label{hegatri3}
   \end{figure}
 An analogous argument holds for antiferromagnetic next-to-nearest neighbor interaction,
 for which we can choose
 \begin{eqnarray*}\nonumber
\phi(\{u^j\}_j)&=&u^{(0,0)}u^{(1,0)}+ u^{(1,0)}u^{(1/2,\sqrt3/2)}+u^{(1/2,\sqrt3/2)}u^{(0,0)}+1\\
&&+c\Bigl(u^{(0,0)}u^{(0,\sqrt  3)}+u^{(0,0)}u^{(3/2,-\sqrt  3/2)}+ 
u^{(0,0)}u^{(-3/2,-\sqrt  3/2)}+1\Bigr).
\end{eqnarray*}
In this case, the six ground states have a striped pattern and are determined by 
their values on two adjacent triangles (see Fig.~\ref{hegatri3}).
\end{example}

\section{Analysis of inhomogeneous energies}\label{inho}
Theorem  \ref{teo:Gamma} can be applied to treat inhomogeneous energies, when the energy
densities themselves depend on the index $i$; namely, we have
\begin{equation}\label{inoF}
F(u,\Omega):=\sum_{i\in\Z^n\cap \Omega}\phi_i(\{u^{i+j}\}_{j\in \Z^n}),
\end{equation}
and $i\mapsto \phi_i:(X)^{\Z^n}\to[0,L]$ is itself $h$-periodic,
and $F_\e$ are defined accordingly as in \eqref{def:en_latt}. Using the approach of Section \ref{simply} such energies can be rewritten as homogeneous energies on a coarser lattice.
Indeed in the notation of that section, it suffices to define
\begin{equation}\label{phon}
\phi(\{\hat u^{\hat\j}\}_{\hat\j})={1\over h^n} \sum_{i\in\{1,\ldots,h\}^n}
\phi_i(\{\Psi(\hat u)^{i+j}\}_{j\in \Z^n}).
\end{equation}
Under the assumption that (up to addition of a constant in order to normalize the minimum to $0$, if necessary) $\phi$ satisfies (H1)--(H3), we can then apply Theorem  \ref{teo:Gamma}. Note that the homogenized energy density $\varphi$ is described by the same formulas
with $F$ as in \eqref{inoF}.

\begin{remark}[general lattices]\rm The extension to inhomogeneous interactions allows to
consider also more general non-Bravais lattices, which are not invariant under translations 
by their elements, e.g., the hexagonal lattice, upon interpreting $\phi_i(\{u^{i+j}\}_{j\in {\cal L}-i})$
as defined on the lattice ${\cal L}-i$. 
\end{remark}

\subsection{Examples: mixtures of spin interactions}
The extension to inhomogeneous energies allows us to consider some homogenization problems
for mixtures of ferromagnetic and anti-ferromagnetic interactions.

\begin{example}[non-frustrated systems]\rm\rm
A simple criterion that ensures that a periodic system of mixtures of ferromagnetic and antiferromagnetic nearest-neighbor interactions in the square lattice is
non-frustrated is that each square has an even number of ferromagnetic interactions.
In this case, if $T$ denotes the period of the system, the value of $u^{(0,0)}$ 
determines the whole non-frustrated $u$, which is $T$-periodic if $u^{(T,0)}=u^{(0,T)}=u^{(0,0)}$,
and it is $2T$ periodic otherwise. Hence, (H1)--(H3) are satisfied with $\phi_i$ given simply 
by the sum of the interactions with one vertex in $i$ (with $(u^j-u^0)^2$ if the interaction between 
$i$ and $i+j$ is ferromagnetic, and $(u^j+u^0)^2$ if it is antiferromagnetic), $K=2$, $h=2T$ and $M'=2M=2h$. Clearly, a non-frustrated system is equivalent to a ferromagnetic system up to a suitable change of sign of, e.g., all sites that take the value $-1$ at a ground state. Hence, the
homogenized energy density is $8\|\nu\|_1$.

A suitable choice of the geometry allows us to exhibit mixtures for which ground states are non-frustrated. We include three two-dimensional examples, with different textures for the minimizers.

(1) We consider a nearest-neighbor interaction system with 
$$
\phi_i(\{u^j\}_j) = \begin{cases} (u^{(1,0)}+u^{(0,0)})^2+  (u^{(0,1)}-u^{(0,0)})^2 & \hbox{ if } i_1 \hbox{ is even}\\
 (u^{(1,0)}-u^{(0,0)})^2+  (u^{(0,1)}-u^{(0,0)})^2 & \hbox{ if } i_1 \hbox{ is odd};
 \end{cases}
 $$
i.e., a two-dimensional system with all ferromagnetic vertical interaction, and alternating columns of antiferromagnetic and ferromagnetic horizontal interactions.  \begin{figure}[h!]
\centerline{\includegraphics [width=1in]{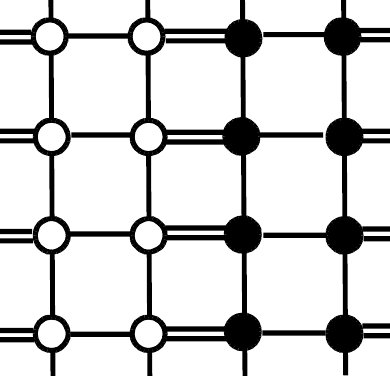}}
\caption{non-frustrated lines}\label{nf1}
   \end{figure}
   The two ground states are then 
alternating vertical pairs of lines of $+1$ and $-1$ spins (see Fig.~\ref{nf1});

(2) We consider a nearest-neighbor interaction system with 
$$
\phi_i(\{u^j\}_j) = (u^{(1,0)}+(-1)^{i_2}\, u^{(0,0)})^2+  (u^{(0,1)}+(-1)^{i_1}\, u^{(0,0)})^2;
$$
i.e., a two-dimensional system with alternating horizontal and vertical lines of ferromagnetic and antiferromagnetic interactions.  \begin{figure}[h!]
\centerline{\includegraphics [width=1.1in]{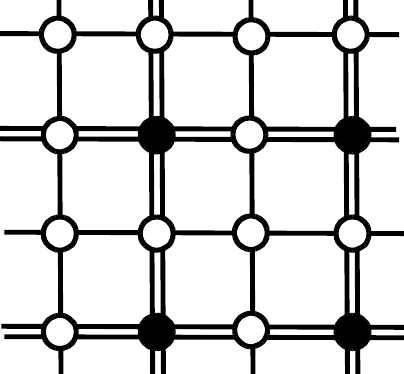}}
\caption{non-frustrated inclusions}\label{nf2}
   \end{figure}
   The two ground states are then uniform states with the opposite value on the lattice $2\ZZ^2$ (see Fig.~\ref{nf2}).

   3) We consider a nearest-neighbor interaction system with 
$$
\phi_i(\{u^j\}_j) = (u^{(1,0)}+(-1)^{i_1+i_2}\, u^{(0,0)})^2+  (u^{(0,1)}+(-1)^{i_1+i_2}\, u^{(0,0)})^2;
$$
i.e., a two-dimensional system with a zig-zag pattern of alternated ferromagnetic and antiferromagnetic interactions.  \begin{figure}[h!]
\centerline{\includegraphics [width=1.1in]{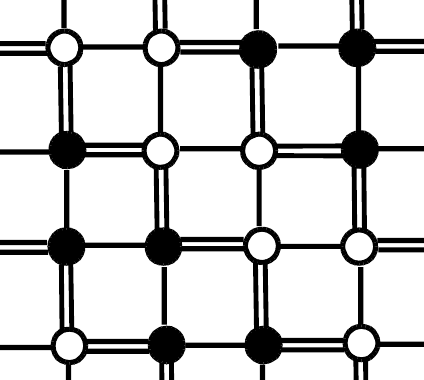}}
\caption{non-frustrated stripes}\label{nf3}
   \end{figure}
   The two ground states are then stripes at $-45$ degrees
(see Fig.~\ref{nf3}).
\end{example}

\begin{remark}[systems parameterized by the `majority phase']\rm
We may consider a $h$-periodic system of nearest-neighbor interactions where the 
antiferromagnetic interactions form a periodic set composed of isolated components.
If the distance between such connected components is large enough then ground state
are non-constant only on isolated sets, so that we may take this `majority value' $\pm1$
as parameterizing the ground states, ensuring a relevant continuum parameter of ferromagnetic type. The $\Gamma$-convergence can be
given in terms of that parameter \cite{BPiatn}. Note however that Theorem \ref{teo:Gamma}
may not directly apply if we have more than one pattern minimizing the isolated
antiferromagnetic contributions, since in this case (H2) would be violated. \end{remark}

The observation above suggests that Theorem \ref{teo:Gamma} may be extended to the case when (H2) does not hold, provided we can define an equivalence class among ground states that can be mixed with zero interfacial energy. The following example clarifies this observation.

\begin{example}[equivalent ground states and interfacial microstructure]\label{equigs}\rm
We consider the three-periodic two dimensional system with
$$
\phi_i(\{u^j\}_j)= \begin{cases} (u^{(1,0)}+u^{(0,0)})^2+  (u^{(0,1)}+u^{(0,0)})^2 & \hbox{ if } i=(0,0) 
\\
 (u^{(1,0)}-u^{(0,0)})^2+  (u^{(0,1)}-u^{(0,0)})^2 & \hbox{  otherwise}.
 \end{cases}
 $$
Note that the system is frustrated; hence, no function satisfies $F(u)=0$.
In this case $\phi$ directly defined as in \eqref{phon} does not satisfy the hypotheses of 
Theorem \ref{teo:Gamma} since its pointwise minimization does not correspond to a ground state
for $F$. We then have to regroup the terms in the sum giving $F$ and renormalize the energy. \begin{figure}[h!]
\centerline{\includegraphics [width=2.5in]{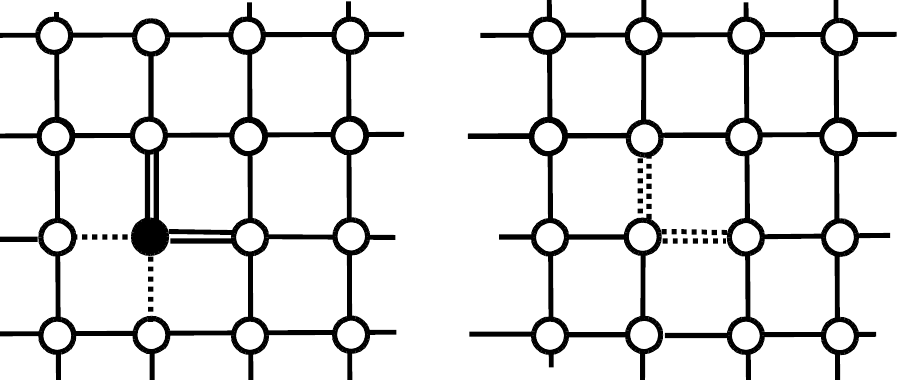}}
\caption{two equivalent ground states}\label{doublemin}
\end{figure}
We can choose $\phi$ defined on functions defined on $3\ZZ^2$ by
\begin{eqnarray*}
\phi(\{\hat u^{\hat\j}\}_{\hat\j})&=&4u^{(1,0)}u^{(0,0)}+4u^{(-1,0)}u^{(0,0)}-8\\
&& + {1\over 2}\sum_{i\in\{-1,0,1,2\}^2} \Bigl((u^{i+e_1}-u^{i})^2+  (u^{i+e_1+e_2}-u^{i+e_1})^2
\\
&& + (u^{i+e_2}-u^{i+e_1+e_2)})^2+  (u^{i}-u^{i+e_2})^2\Bigr),
\end{eqnarray*}
where $u=\Psi(\hat u)$. 
Minimizing this energy density we obtain two pairs of functions defined on
$\{-1,0,1,2\}^2$ which have a constant value on all points except at $(0,0)$ where they can take equivalently the value $+1$ or $-1$. In particular, two functions with the same constant value on most points give two `equivalent' ground states, with no interfacial energy between them (see Fig.~\ref{doublemin}).

 \begin{figure}[h!]
\centerline{\includegraphics [width=1.8in]{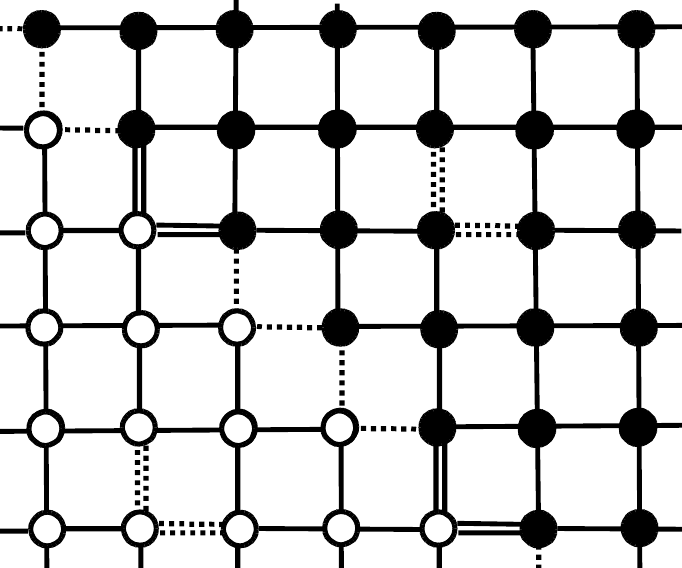}}
\caption{optimal interfacial microstructure}\label{doubleminint}
\end{figure}
\end{example}
Note that the optimal microstructure for the interfacial energy is obtained by 
maximizing non-frustrated interactions along the interface (see  e.g.~Fig.~\ref{doubleminint} for an optimal microstructure for a 45-degree interface).

\begin{example}[a non-coercive case]\label{nonco}\rm 
We show an example of application of Theorem \ref{teo:Gamma} to a 
case when (H1) and (H2) are not satisfied. As a consequence, the resulting 
$\Gamma$-limit is degenerate and non-coercive on partitions of sets of finite 
perimeter.

We consider the inhomogeneous system on $\ZZ^2$ with
\begin{eqnarray*}
\phi_i(\{u^j\}_j)=\begin{cases}
(u^{(1,0)}-u^{(0,0)})^2+  (u^{(1,1)}-u^{(1,0)})^2
\\
+ (u^{(0,1)}-u^{(1,1)})^2+  (u^{(0,0)}-u^{(0,1)})^2  &\hbox{ if } i_1=0 \hbox{ modulo }4\\
(u^{(1,0)}-u^{(0,0)})^2+  (u^{(1,1)}+u^{(1,0)})^2
\\
+ (u^{(0,1)}-u^{(1,1)})^2+  (u^{(0,0)}-u^{(0,1)})^2-1  &\hbox{ if } i_1= 1 \hbox{ modulo }4\\
(u^{(1,0)}+u^{(0,0)})^2+  (u^{(1,1)}+u^{(1,0)})^2
\\
+ (u^{(0,1)}+u^{(1,1)})^2+  (u^{(0,0)}+u^{(0,1)})^2  &\hbox{ if } i_1= 2 \hbox{ modulo }4\\
(u^{(1,0)}+u^{(0,0)})^2+  (u^{(1,1)}-u^{(1,0)})^2
\\
+ (u^{(0,1)}+u^{(1,1)})^2+  (u^{(0,0)}+u^{(0,1)})^2-1  &\hbox{ if } i_1= 3 \hbox{ modulo }4.
\end{cases}
\end{eqnarray*}
These interactions are four periodic in the horizontal direction, are ferromagnetic on 
squares with the left-hand side vertices $i$ with $i_1=0$ modulo $4$ and antiferromagnetic
on squares with the left-hand side vertices $i$ with $i_1=2$ modulo $4$. On the other squares
the horizontal interactions are the same as the vertical interaction on the left edge, so that
we have three ferromagnetic interactions and a antiferromagnetic one or the converse,
which are normalized accordingly. 

 \begin{figure}[h!]
\centerline{\includegraphics [width=2.8in]{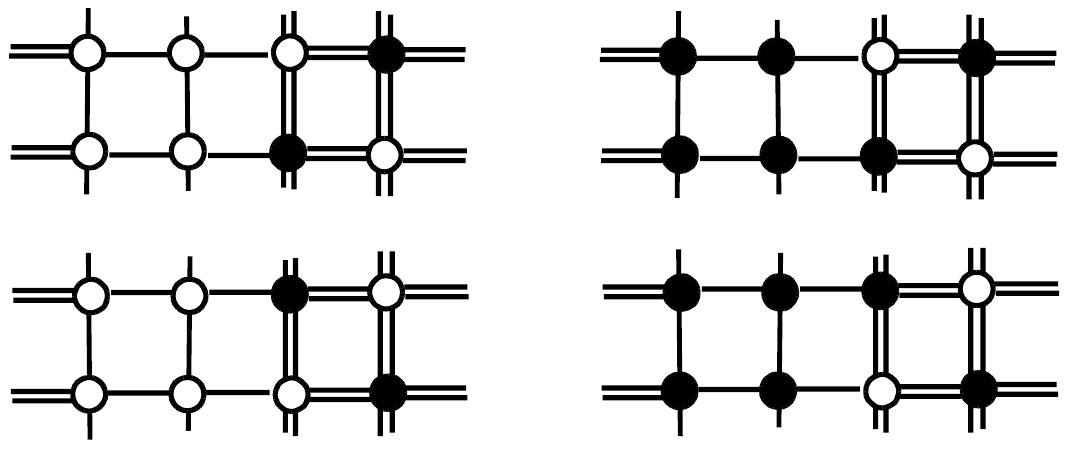}}
\caption{minimal patterns}\label{degenerate4}
\end{figure}
We note that functions with zero energy restricted to four-by-two cells have one of the four patterns in Fig.~\ref{degenerate4}.
Each of these functions $u$ can be viewed as a pair of a uniform ferromagnetic state and an alternating antiferromagnetic state, so that it can be identified, after extension by periodicity, with the value $(u^{(0,0)}, u^{(2,0)})\in \{1,-1\}^2$.

 \begin{figure}[h!]
\centerline{\includegraphics [width=2.2in]{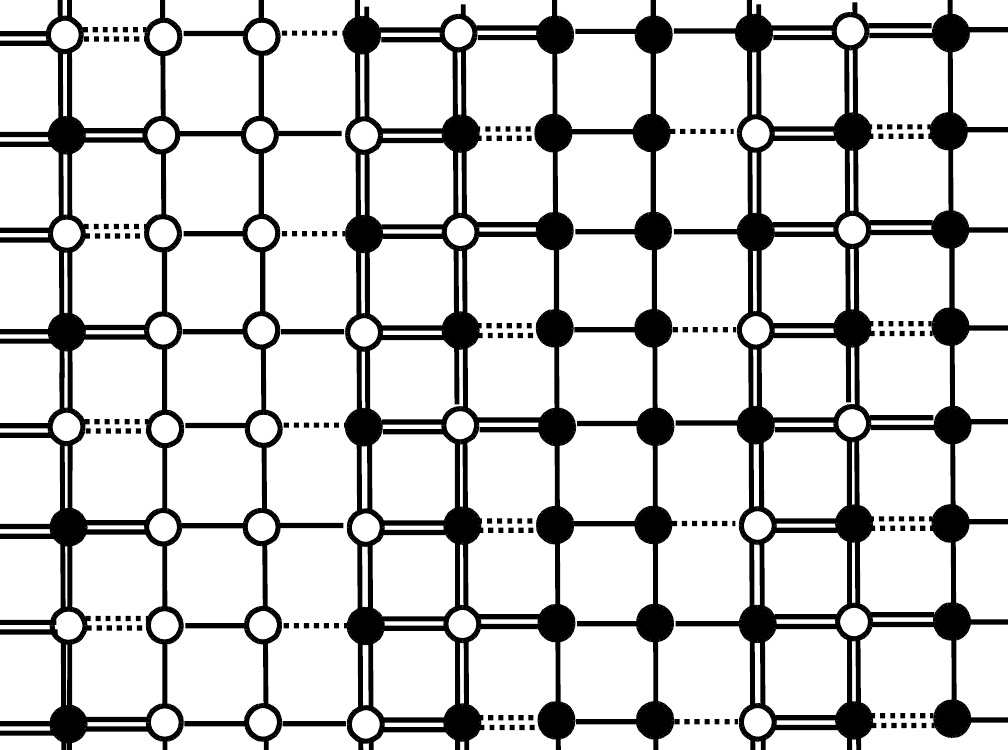}}
\caption{a vertical interface with zero energy}\label{degor}
\end{figure}
We can apply Theorem \ref{teo:Gamma} using $l\in\{1,-1\}^2$ as parameters, with a slight abuse of notation. 
Note that the energy density $\varphi$ is degenerate on vertical interfaces; i.e.,
$$
\varphi(l,l',e_1)=0 \hbox{ for all } l,l'\in  \{1,-1\}^2
$$
(see Fig.~\ref{degor} for a vertical interface between $(1,1)$ and $(-1,1)$ with zero energy), 
while for horizontal interfaces we have
$$
\varphi(l,l',e_1)={1\over 2} |l-l'|^2.
$$
 \begin{figure}[h!]
\centerline{\includegraphics [width=2.8in]{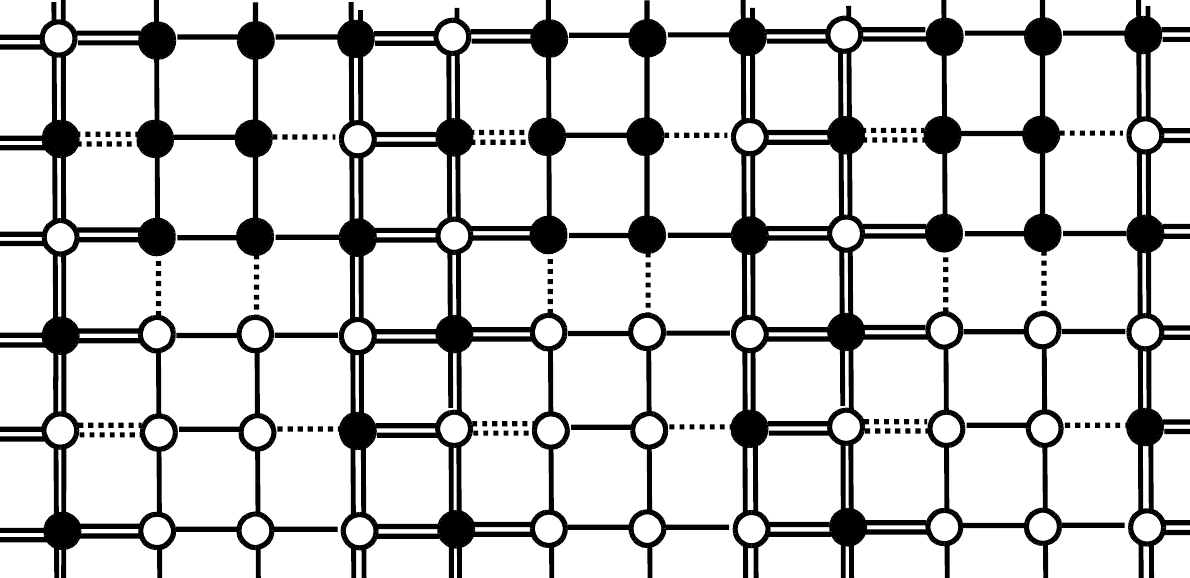}}
\caption{a horizontal minimal interface}\label{degint}
\end{figure}An example of minimal horizontal interface between $(1,1)$ and $(-1,1)$ is given in Fig.~\ref{degint}.
By approximation of an arbitrary interface with interfaces in the coordinate directions 
we obtain 
$$
\varphi(l,l',\nu)={1\over 2} |l-l'|^2|\nu_1|.
$$

Note that, since our system does not satisfy (H1) and (H2),  the compactness Theorem \ref{teo:comp} cannot be applied.

\end{example}

\begin{example}[some degenerate cases] 
\rm We show some cases when we cannot apply 
 Theorem \ref{teo:Gamma}. 
 
(i) We consider a system as in the Example \ref{equigs} but with a shorter periodicity; namely,
with $\phi$ defined on functions defined on $2\ZZ^2$ by
\begin{eqnarray*}
\phi(\{\hat u^{\hat\j}\}_{\hat\j})&=&(u^{(1,0)}+u^{(0,0)})^2+  (u^{(0,1)}+u^{(0,0)})^2-8
\\
&& + (u^{(-1,0)}-u^{(0,0)})^2+  (u^{(0,-1)}-u^{(0,0)})^2 
\\
&& + {1\over 2}\Bigl( (u^{(1,0)}-u^{(1,1)})^2+  (u^{(1,1)}-u^{(0,1)})^2
\\
&& + (u^{(0,1)}-u^{(-1,1)})^2+  (u^{(-1,1)}-u^{(-1,0)})^2
\\
&& + (u^{(-1,0)}-u^{(-1,-1)})^2+  (u^{(-1,-1)}-u^{(0,-1)})^2
\\
&& + (u^{(0,-1)}-u^{(1,-1)})^2+  (u^{(1,-1)}-u^{(0,0)})^2\Bigr),
\end{eqnarray*}
where $u=\Psi(\hat u)$.  \begin{figure}[h!]
\centerline{\includegraphics [width=2.5in]{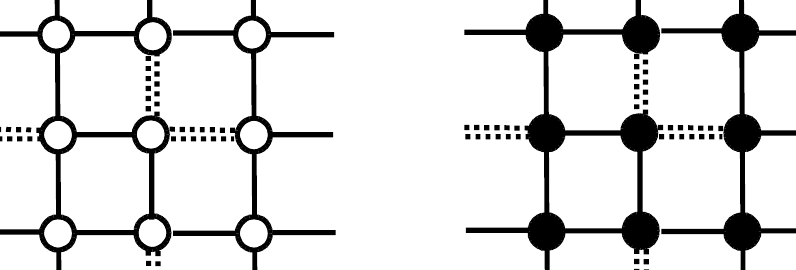}}
\caption{uniform minimal states with highlighted frustrated interactions}\label{frust}
\end{figure}
In this case we have non-periodic minimal states, since 
we can construct 45-degree interfaces with zero interfacial energy between two uniform minimal periodic states 
as those in Fig.~\ref{frust}.
 \begin{figure}[h!]
\centerline{\includegraphics [width=1.6in]{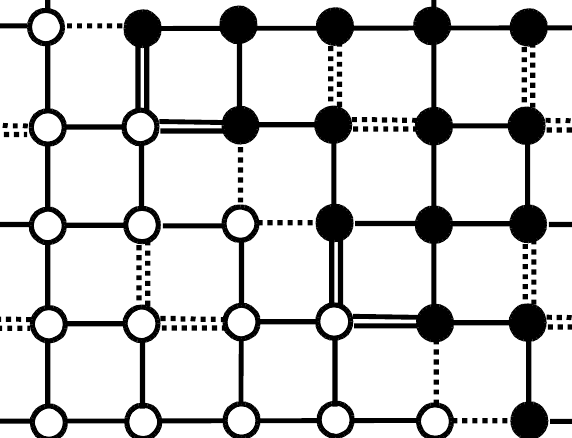}}
\caption{interface with zero energy}\label{minni}
\end{figure}
 Hence, we can construct alternating layers of those two states 
of arbitrary (non-periodic) thickness (see  Fig.~\ref{minni}).
Note that in this example the minimum of $\phi$ is negative, even though $F\ge0$. 

This examples mixes the features of Example \ref{equigs} (equivalent ground states)
and Example \ref{nonco} (degeneracy in one directions).

\smallskip
(ii) (a totally frustrated example). We now give a two-dimensional example 
on the square lattice with a behavior similar to antiferromagnetic interactions on a triangular lattice. In this example the interactions corresponding to the sides of a unit lattice square are three ferromagnetic interactions and one antiferromagnetic interaction, or the converse. We choose 
the interactions so that all squares in a horizontal line are of the same type.
Since minimizers have only one of the four interactions frustrated, interactions are easily normalized. We may choose $\phi_i$ as follows:
$$
\phi_i(\{u^j\}_j)=\begin{cases}
(u^{(1,0)}+u^{(0,0)})^2+  (u^{(1,1)}-u^{(1,0)})^2
\\
\quad+ (u^{(0,1)}-u^{(1,1)})^2+  (u^{(0,0)}-u^{(0,1)})^2-4
& \hbox{ if $i_2$ is even}
\\
(u^{(1,0)}-u^{(0,0)})^2+  (u^{(1,1)}+u^{(1,0)})^2
\\
\quad+ (u^{(0,1)}+u^{(1,1)})^2+  (u^{(0,0)}+u^{(0,1)})^2-4
& \hbox{ if $i_2$ is odd}
\end{cases}
$$
Particular ground states of this energy are the functions $v$ and $-v$, where
$v^i=(-1)^{\lfloor i_2/2\rfloor}$ (double alternate horizontal stripes). 

 \begin{figure}[h!]
\centerline{\includegraphics [width=1.6in]{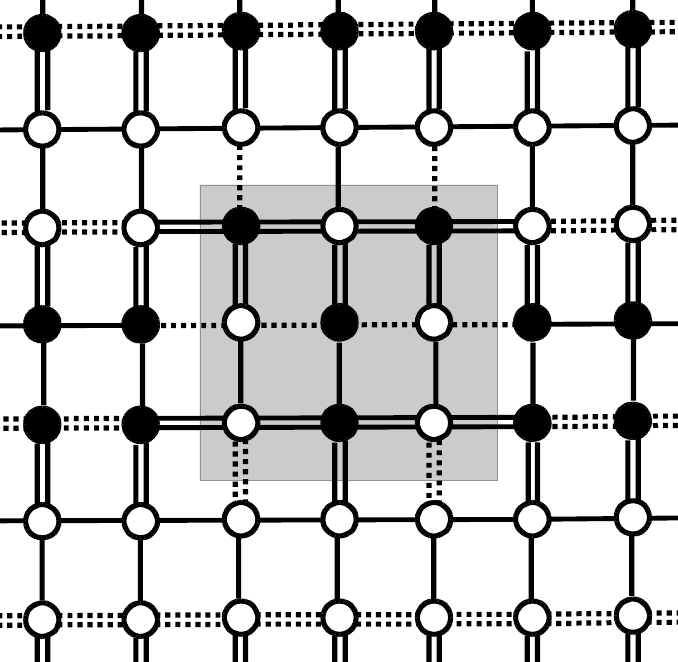}}
\caption{inclusion with no interfacial energy}\label{frusq}
\end{figure}
In order to show that the energy is not minimized only on periodic ground states it suffices to
exhibit a function with $F(u)=0$ consisting of a compact-support perturbation of the function $v$ defined above. Such a function $u$ is described in Fig.~\ref{frusq}.
\end{example}

\section{Conclusions}
We have provided a characterization of lattice systems with interfacial energies between modulated phases and textures. The result relies on the possibility of identifying a finite number of ground states that completely describe locally the possible patterns. These ground states themselves play the role of the parameters in the limit description, that hence can be set in a (much) higher-dimensional space. The overall behavior of the interfaces has been described 
by homogenization formulas that account for surface microstructure. We noted that such
microstructure is not a characteristic of the ground states only but depends on 
the details of the interactions.
 
We have shown a number of spin systems which can be described by our result, but also 
some frustrated systems for which we have infinitely many periodic and non-periodic ground states.
For some of such systems we can still pinpoint a finite number of ground states locally characterizing functions with bounded energy and some weak form of coerciveness holds
that may make a (partial) description using our result still meaningful.  Nevertheless, we also exhibited `totally frustrated' systems for which no strong or weak type of interfacial energy 
seems to make sense. The problem remains open of distinguishing between the 
types of situations not covered by our results.

\bibliographystyle{plain}
\baselineskip=11pt
\bibliography{bib-BC}

\begin{thebibliography}{10}

\bibitem{ABC-libro}
R.~Alicandro, Braides A., and M.~Cicalese.
\newblock Book in preparation.

\bibitem{ABC}
R.~Alicandro, A.~Braides, and M.~Cicalese.
\newblock Phase and anti-phase boundaries in binary discrete systems: a
  variational viewpoint.
\newblock {\em Netw. Heterog. Media}, 1(1):85--107, 2006.

\bibitem{ACR}
R.~Alicandro, M.~Cicalese, and M.~Ruf.
\newblock Domain formation in magnetic polymer composites: an approach via
  stochastic homogenization.
\newblock {\em Arch. Rat. Mech. Anal.}, 218(2):945--984, 2015.

\bibitem{AlGe}
R.~Alicandro and M.~S. Gelli.
\newblock Local and non local continuum limits of {I}sing type energies for
  spin systems.
\newblock {\em Preprint}, 2014.

\bibitem{AmBrI}
L.~Ambrosio and A.~Braides.
\newblock Functionals defined on partitions in sets of finite perimeter, {I}:
  Integral representation and gamma-convergence.
\newblock {\em J. Math. Pures Appliquees}, 69:285--306, 1990.

\bibitem{AmBrII}
L.~Ambrosio and A.~Braides.
\newblock Functionals defined on partitions of sets of finite perimeter, {II}:
  semicontinuity, relaxation and homogenization.
\newblock {\em J. Math. Pures Appl}, 69:307--333, 1990.

\bibitem{BBR}
G.~Bellettini, A.~Braides, and G.~Riey.
\newblock Variational approximation of aniso\-tropic functionals on partitions.
\newblock {\em Annali di Matematica Pura ed Applicata}, 184(1):75--93, 2005.

\bibitem{GCB}
A.~Braides.
\newblock {\em {$\Gamma$}-convergence for beginners}, volume~22 of {\em Oxford
  Lecture Series in Mathematics and its Applications}.
\newblock Oxford University Press, Oxford, 2002.

\bibitem{ICM}
A.~Braides.
\newblock Discrete-to-continuum variational methods for lattice systems.
\newblock {\em Proceedings International Congress of Mathematicians. Seoul}, 2014,
  pages 997--1015.

\bibitem{B-planelike}
A.~Braides.
\newblock An example of non-existence of plane-like minimizers for an
  almost-periodic {I}sing system.
\newblock {\em Journal of Statistical Physics}, 157(2):295--302, 2014.

\bibitem{BCP}
A.~Braides and V.~Chiad{\`o}~Piat.
\newblock Integral representation results for functionals defined on {$SBV
  (\Omega; R^m)$}.
\newblock {\em Journal de Math{\'e}matiques Pures et Appliqu{\'e}es},
  75(6):595--626, 1996.

\bibitem{BCPS}
A.~Braides, V.~Chiad\`o Piat, and M.~Solci.
\newblock Discrete double-porosity models for spin systems.
\newblock {\em Mathematics and Mechanics of Complex Systems}, to appear.

\bibitem{HMI}
A.~Braides and A.~Defranceschi.
\newblock {\em Homogenization of multiple integrals}.
\newblock Oxford University Press, Oxford, 1998.

\bibitem{BGP}
A.~Braides, A.~Garroni, and M.P. Palombaro.
\newblock Interfacial energies of systems of chiral molecules.
\newblock {\em Preprint 2015}.

\bibitem{BPiatn2}
A.~Braides and A.~Piatnitski.
\newblock Variational problems with percolation: dilute spin systems at zero
  temperature.
\newblock {\em Journal of Statistical Physics}, 149(5):846--864, 2012.

\bibitem{BPiatn}
A.~Braides and A.~Piatnitski.
\newblock Homogenization of surface and length energies for spin systems.
\newblock {\em Journal of Functional Analysis}, 264(6):1296--1328, 2013.

\bibitem{CDLL}
L.A. Caffarelli and R.~de~la Llave.
\newblock Planelike minimizers in periodic media.
\newblock {\em Communications on Pure and Applied Mathematics},
  54(12):1403--1441, 2001.

\bibitem{CHS}
A.~Ciach, J.~S. Hoye, and G.~Stell.
\newblock Microscopic model for microemulsions.
\newblock {\em Journal of Physics A: Mathematical and General}, 21(15):L777,
  1988.

\bibitem{CMM}
B.~D. Coleman, M.~Marcus, and V.~J. Mizel.
\newblock On the thermodynamics of periodic phases.
\newblock {\em Archive for Rational Mechanics and Analysis}, 117(4):321--347,
  1992.

\bibitem{DGL}
E.~De~Giorgi and G.~Letta.
\newblock Une notion g{\'e}n{\'e}rale de convergence faible pour des fonctions
  croissantes d'ensemble.
\newblock {\em Annali della Scuola Normale Superiore di Pisa-Classe di
  Scienze}, 4(1):61--99, 1977.

\bibitem{GLL}
A.~Giuliani, J.~L. Lebowitz, and E.~H. Lieb.
\newblock {I}sing models with long-range antiferromagnetic and short-range
  ferromagnetic interactions.
\newblock {\em Physical Review B}, 74(6):064420, 2006.

\bibitem{GLL3}
A.~Giuliani, J.~L. Lebowitz, and E.~H. Lieb.
\newblock Checkerboards, stripes, and corner energies in spin models with
  competing interactions.
\newblock {\em Physical Review B}, 84(6):064205, 2011.

\bibitem{GLS}
A.~Giuliani, E.~H. Lieb, and R.~Seiringer.
\newblock Realization of stripes and slabs in two and three dimensions.
\newblock {\em Physical Review B}, 88(6):064401, 2013.

\bibitem{OK}
T.~Ohta and K.~Kawasaki.
\newblock Equilibrium morphology of block copolymer melts.
\newblock {\em Macromolecules}, 19(10):2621--2632, 1986.

\bibitem{SA}
M.~Seul and D.~Andelman.
\newblock Domain shapes and patterns: the phenomenology of modulated phases.
\newblock {\em Science}, 267(5197):476--483, 1995.

\end{thebibliography}

\bigskip \noindent{\sc Authors' addresses}

\medskip
{\parindent=0pt\baselineskip=10.5pt

{\sc Andrea Braides}

Dipartimento di Matematica, Universit\`a di Roma `Tor Vergata'

via della Ricerca Scientifica, 00133 Roma, Italy

e-mail: braides@mat.uniroma2.it

\smallskip

{\sc Marco Cicalese}

Zentrum Mathematik - M7, Technische Universit\"at M\"unchen

Boltzmannstrasse~3, 85748 Garching, Germany

e-mail: cicalese@ma.tum.de}
\end{document}